\documentclass[10pt,journal,compsoc]{IEEEtran}
\usepackage{amsmath}
\usepackage{mathtools}
\usepackage{afterpage}
\usepackage{wrapfig}

\ifCLASSOPTIONcompsoc
  \usepackage[nocompress]{cite}
\else
  \usepackage{cite}
\fi
\ifCLASSINFOpdf
  \usepackage[pdftex]{graphicx}
\else
\fi
\begin{document}
%

\title{Read-Tuned STT-RAM and eDRAM Cache Hierarchies for Throughput and Energy Enhancement}

%
%
%
%

\author{Navid~Khoshavi,
	    Xunchao~Chen,
	     Jun~Wang,~\IEEEmembership{Senior Member,~IEEE,}
               and~Ronald~F.~DeMara,~\IEEEmembership{Senior Member,~IEEE}
\IEEEcompsocitemizethanks{\IEEEcompsocthanksitem N. Khoshavi, X. Chen, J. Wang, and R.F. DeMara are with the Department
of Electrical and Computer Engineering, University of Central Florida, Orlando,
FL, 32816.\protect\\
E-mail: see http://cal.ucf.edu}
\thanks{Manuscript received xx, 2016; revised TBD xx, 2016.}}

\maketitle


%
\IEEEpeerreviewmaketitle

%
%
%
%

\begin{abstract}


As capacity and complexity of on-chip cache memory hierarchy increases, the service cost to the critical loads from Last Level Cache (LLC), which are frequently repeated, has become a major concern. The processor may stall for a considerable interval while waiting to access the data stored in the cache blocks in LLC, if there are no independent instructions to execute. To provide accelerated service to the critical loads requests from LLC, this work concentrates on leveraging the additional capacity offered by replacing SRAM-based L2 with Spin-Transfer Torque Random Access Memory (STT-RAM) to accommodate frequently accessed cache blocks in exclusive read mode in favor of reducing the overall read service time. Our proposed technique partitions L2 cache into two STT-RAM arrangements with different write performance and data retention time. The retention-relaxed STT-RAM arrays are utilized to effectively deal with the regular L2 cache requests while the high retention STT-RAM arrays in L2 are selected for maintaining repeatedly read accessed cache blocks from LLC by incurring negligible energy consumption for data retention. Our experimental results show that the proposed technique can reduce the mean L2 read miss ratio by 51.4\% and increase the IPC by 11.7\% on average across PARSEC benchmark suite while significantly decreasing the total L2 energy consumption compared to conventional SRAM-based L2 design. 


\end{abstract}


\section{Introduction}\label{sec:introduction}

\IEEEPARstart{T}{he} increasing bandwidth demand of current memory-intensive applications incurs significant data movement that negatively impacts off-chip bandwidth, on-chip memory access latency, and energy consumption \cite{borkar-dac-2007} \cite{kurian-ics-2014} \cite{chen_dac_2016} \cite{ahmadian-2014}. To reduce data transfer between on-chip and off-chip memory components, commercial multi-core systems utilize multi-level cache methodology \cite{borkar-acm-2011} \cite{TMSCS-memory} \cite{Imani-GLSVLSI-2016} whereby fast, low-capacity, and high leakage power SRAM arrays are employed in the upper-levels of cache, i.e. L1 and L2, while large, low leakage power and high refresh demand eDRAM is placed in LLC. The employment of relatively spacious SRAM arrays as L2 cache design in the middle of cache hierarchy results in two major challenges: 1) \textit{leakage}: the high leakage power characteristic of SRAM cells results in excessive power budget \cite{jog-dac-2012}, and 2)  \textit{area}: capacity constraints induced by the SRAM cell footprint prevent favorable residency of the working set to reside close to the active core \cite{jaleel-hpca-2006} \cite{Imani-NVMSA-2016}. The negative effect of aforementioned issues is exacerbated with the high dynamic power dissipation incurred to drive on-chip interconnects while exchanging data \cite{bojnordi2013desc}\cite{udipi2009non}.  


While other recent works have made significant advancement to optimize eDRAM LLC (L3 in this work) \cite{loh-micro-2011} \cite{arroyo-rd-2011}, in this paper we concentrate on applying new insights regarding working set behavior to optimize L2 cache using a heterogeneous STT-RAM. STT-RAM offers a promising alternative solution to take the place of area-inefficient and high leakage power SRAM technology. STT-RAM is well-known for its non-volatility and near-zero standby power while offering at least 3x to 4x area savings compared to SRAM \cite{chang-hpca-2013} \cite{smullen-hpca-2011}. 
This work is an initiative study to answer how much performance gain and energy reduction can be achieved by replacing the conventional SRAM-based L2 with STT-RAM. To maximize the benefit of extra capacity realized by this replacement, we have proposed a novel technique called \textit{Read Reference Activity Persistent (RRAP)} strategy that accelerates the service to critical requests while also efficiently manages regular L2 cache requests. Essentially, the loads and stores exhibit different levels of criticality in the processor. While the stores requests can be postponed through buffering before cache or memory commitment, the loads requests demand immediate treatment to maintain the stalls within an acceptable delay period in pipelined processors \cite{khan2014improving}.  

However, the cache memory shortage to maintain continuously expanding working sets close to the core causes the performance degradation or increased miss ratio. This degradation is exacerbated by introducing more cache levels with larger LLC capacity. For instance, the transfer latency for a cache block from L2 to L1 in hyper-threaded Pentium IV is 18 cycles, while this latency goes up to 360 cycles for data movement between main memory and L2 \cite{tuck2003initial}.
Thus, the processor may stall for a long time to access the data stored in the cache blocks in LLC if there are no independent instructions to execute. 

To provide accelerated service to the critical loads requests from LLC, RRAP technique targets Immense Read Reused Access (IRRA) blocks that remain unchanged to be brought from LLC to hybrid L2 cache design in favor of reducing read miss ratio in L2, which in turn causes the reduction of overall read service time.

\begin{figure*}[t] 
\centering
\includegraphics[width=7.2in, keepaspectratio]{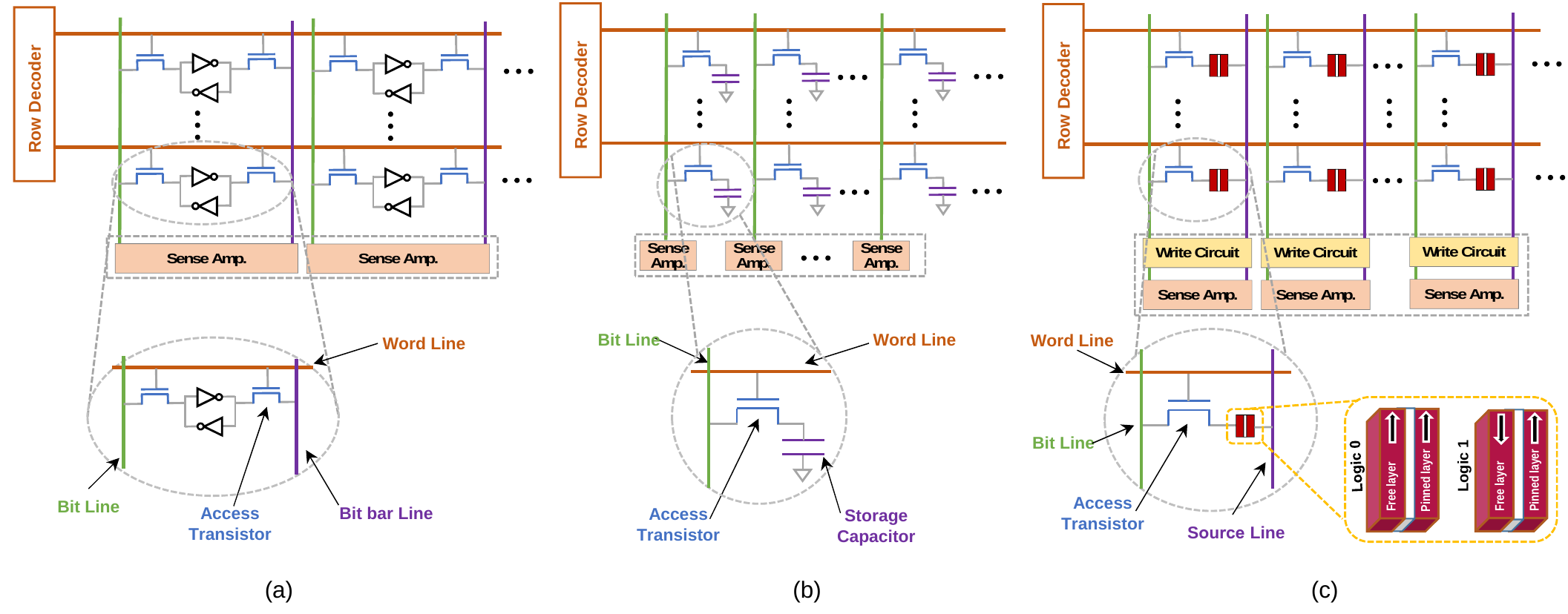}
\vspace*{-0.2in}
\caption{LLC organization based on (a) SRAM, (b) eDRAM, (c) STT-RAM.}
\label{fig:cacheTech}
\end{figure*}

To accomplish this, RRAP initially employs a strategy to exploit the non-volatility of STT-RAM for storing reused lines by read operations. Second, it takes advantage of retention-relaxed STT-RAM having improved write performance to manage regular cache requests while maintaining the high cache utilization offered by SRAM.
Compared to the existing works, this paper provides the following contributions:
\begin{itemize}
\item new insights regarding the distribution of read reused cache blocks within cache for PARSEC 2.1 suite benchmarks,
\item a low-conflict in-situ monitoring mechanism to track LLC traffic and autonomously copying the exclusive read reused lines to upper-level High Retention STT-RAM Cache (HRSC) of each core,
\item improve the overall read service time by accelerating the service to the critical loads versus stores,
\item identify preferred designs for Low Retention STT-RAM Cache (LRSC) configuration by comprehensively evaluating the candidates in terms of energy consumption and performance delivery, and
\item optimization strategies for balancing workload lifetime performance versus energy reduction from diminished write latency.
\end{itemize}

The remainder of the paper is organized as follows:  The technology trends for cache organization are introduced in Section 2. Section 3 identifies the motivation behind the RRAP strategy. Section 4 presents the technical approach and the details of the proposed method. We summarize RRAP experimental results in Section 5. The related works are discussed in Section 6. Section 7 concludes the paper.    

\section{Background on Technology Trends for Cache Organization}
The deployment of large SRAM in favor of the performance improvement comes with the cost of increased energy consumption and high area overhead. To address this issue, the researchers started to look into the alternative solutions which offer the higher energy-efficient, cost-effective, large-capacity, and competitive read/write operation speed compared to the traditional SRAM arrays as illustrated in Fig. \ref{fig:cacheTech}. 

The embedded DRAM (eDRAM) is one of the promising solutions pioneered in the design of on-chip memory hierarchy. As depicted in Fig. \ref{fig:cacheTech} (b), the eDRAM cells are arranged in a two-dimensional array whereby each storage capacitor is connected to the \textit{bitline} wire through the access transistor. Even though the employment of capacitor for maintaining the logic value has significantly saved the area compared to the SRAM utilizing 6 transistors per bit cell, the eDRAM technology suffers from high dynamic energy consumption due to mandatory periodic refresh required to keep the stored value in the valid state \cite{valero2015design}. It has been reported in \cite{agrawal2013refrint} that refresh scheme contributes around 70\% to the overall energy consumption in LLC. Furthermore, the periodic refresh operation limits the CPU accesses to the eDRAM arrays which leads to typically deployment of the eDRAM as LLC by the memory designers.   

Another promising alternative solution to replace with SRAM is STT-RAM. The non-volatility and near-zero leakage power are two prominent characteristics of STT-RAM cell. Furthermore, the STT-RAM offers at least 3x to 4x area savings compared to the SRAM \cite{samavatian2014efficient}. As illustrated in Fig. \ref{fig:cacheTech} (c), the STT-RAM cache leverages an extra device to drive large current for a certain period when the write operation is issued. The high write current and slow write operation of STT-RAM has motivated the researchers to devise the novel architectures to overcome these new challenges while developing STT-RAM based L2 between L1 and LLC \cite{chang-hpca-2013} \cite{sun-micro-2011}. Herein, we address all of the above issues through adjustment of the intrinsic non-volatility characteristics of STT-RAM to achieve SRAM-competitive performance while incurring low energy consumption.

\section{Motivation}\label{sec:motivation}
\begin{figure}[ht] 
\centering
\includegraphics[width=3.3in, keepaspectratio]{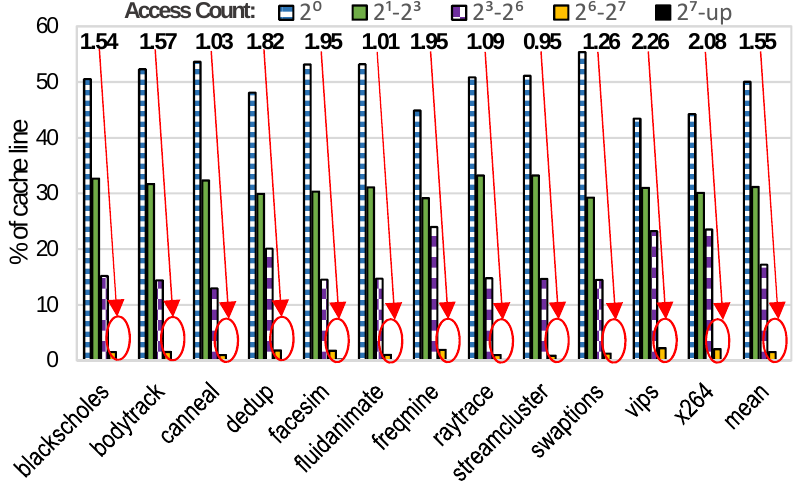}
\vspace*{-0.1in}
\caption{Percentage of total cache lines experiencing read operation.}
\label{fig:cache_line_RRA}
\end{figure}

The cache lines, which are brought into a shared LLC, can be classified into two categories, namely, \textit{non-reused} and \textit{reused} cache lines. A non-reused cache line in LLC does not experience more than one read access, while a reused cache line is accessed from multiple cores during its residency in the LLC. Fig. \ref{fig:cache_line_RRA} illustrates the distribution of read accesses to the cache line, in which the majority of the LLC lines are non-reused cache lines, as indicated by the leftmost blue column for each benchmark which averages 50.1\%  across the benchmark suite \footnote{The experimental setup is explained in Section \ref{sec:experiment}}.

\begin{figure}[ht] 
\centering
\includegraphics[width=3.4in, keepaspectratio]{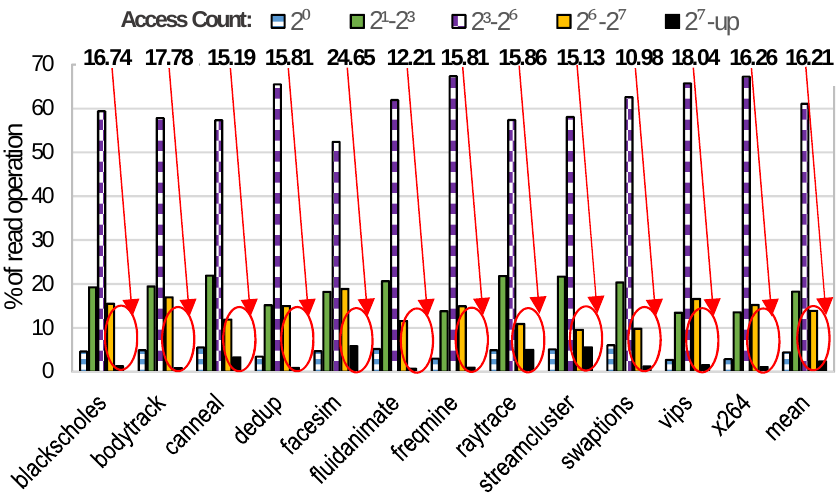}
\vspace*{-0.1in}
\caption{Proportion of read reference activity.}
\label{fig:read_ref}
\end{figure}

Although the majority of the LLC line read accesses are non-reused cache lines, this portion of cache space is accessed only once for read operation purpose during program execution which results in spending a small fraction of time to read from the non-reused cache lines. On the other hand, reused cache lines with high read access rate, e.g. read more than 64 times, have experienced a significant amount of program read operations, e.g. 16.2\% on average, while they only occupy 1.55\% of the entire cache space as depicted in Fig. \ref{fig:cache_line_RRA} and \ref{fig:read_ref}. 

\begin{figure}[ht] 
\centering
\includegraphics[width=3.3in, keepaspectratio]{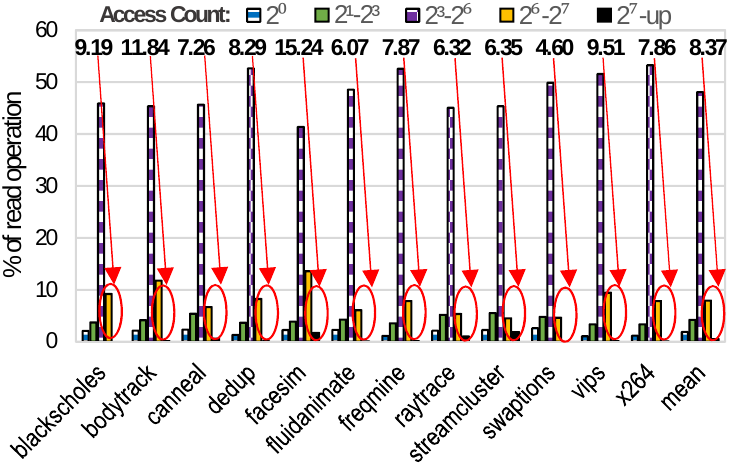}
\vspace*{-0.1in}
\caption{Proportion of exclusive-read reference activity.}
\label{fig:proportion_RRA_0other}
\end{figure}

These reused cache lines can experience either \textit{read-only access} or \textit{multifaceted access,} e.g. write, insert, evict. The exclusive-read reused cache lines, whose read accesses are more than 64, account for 8.37\% of the total read accesses as shown in Fig. \ref{fig:proportion_RRA_0other}. Herein, we refer to this group of cache lines as Immense Read Reused Access (IRRA) blocks that remain unchanged during program runtime. This observation motivated us to re-design the conventional cache design in favor of IRRA cache blocks to reduce the response time to read requests from LLC which in turn cause overall IPC improvement. Moreover, our goal align with the recent works that attempt to prioritize the service to performance critical read reused cache blocks \cite{khan2014improving}\cite{pekhimenko2015exploiting}\cite{seshadri2015mitigating} for achieving a better performance.

%


The IRRA blocks are excellent choices to be stored into high retention STT-RAM arrays because 1) these lines are written once while experiencing read-only operations for the remainder of program execution, 2) the energy and performance overheads for accessing IRRA blocks are small because once the IRRA block is brought to the high retention STT-RAM array, it only will be accessed by the read operations prior to its eviction.  

The characteristics of high versus low retention memory arrays are identified in the following Section.


 \vspace*{-0.2in}


\section{Technical Approach}
\begin{figure}[t] 
\centering
\includegraphics[width=3.5in, keepaspectratio]{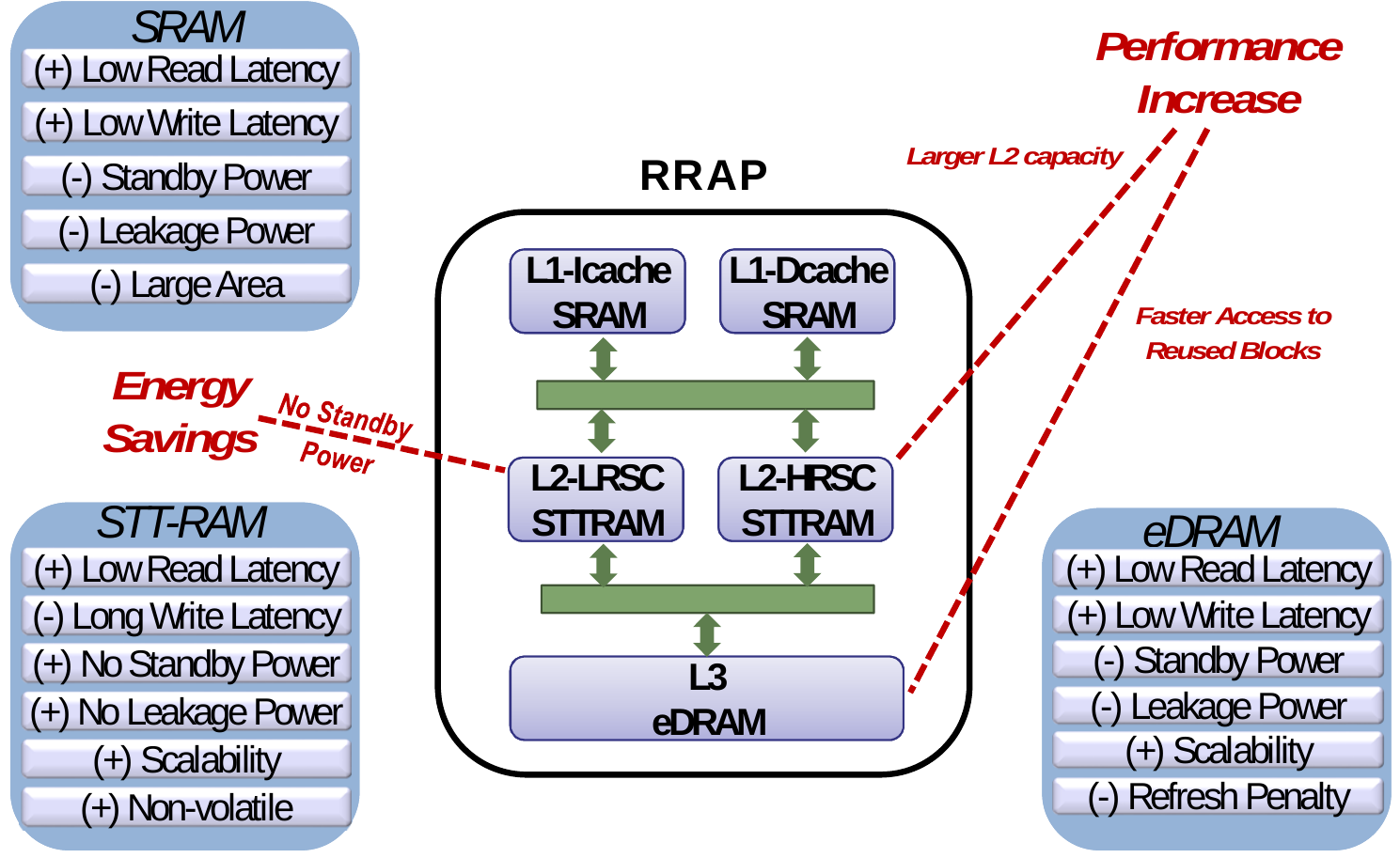}
 \vspace*{-0.1in}
\caption{RRAP heterogeneous split cache architecture.}
\label{fig:proposed_method}
\end{figure}

The intuition behind RRAP is to benefit from the extra capacity provided by replacing SRAM-based L2 with STT-RAM to accommodate the replica of IRRA cache lines from LLC. 
The proposed schematic view is illustrated in Fig. \ref{fig:proposed_method} and referred as Read Reference Activity Persistent (RRAP) cache design, in which the service to frequently-read cache lines of LLC is accelerated while energy consumption is significantly amortized due to near-zero standby power property of STT-RAM arrays. To achieve this goal, the non-inclusive cache allocation policy is adopted in RRAP \cite{TMSCS-noninclusive}. During the program execution, the replica cache block in upper-level of cache is frequently accessed, while the duplicate copy of block in LLC remains unused. By reducing the access to the duplicate block in LLC, this cache block will be eventually selected as a candidate for eviction due to the replacement policy. However, the eviction of this cache block must not back-invalidate the replica cache block in the upper-level of cache. Accordingly, the read service time is not impacted by cache replacement policy in LLC.


In order to effectively manage regular L2 request, the non-volatility of STT-RAM needs to be relaxed through modulating write pulse width and write current. By replacing SRAM with LRSC, the reduced die area provides additional capacity which can accommodate LLC frequent read reference lines with no extra energy consumption for data preservation. We have allocated this extra space to HRSC which retains the data for \textit{10 years}. The combination of LRSC and HRSC in an L2 structure not only provides energy benefit, but also improves performance as will be shown herein.

\subsection{Read Reference Activity Persistent (RRAP) Cache Hierarchy}
Since the non-inclusive policy is adopted in RRAP, the cache blocks are brought into both the LRSC and LLC upon an LLC miss. Upon an LLC write hit, the requested block is brought into LRSC while the replica remains in the LLC. Likewise, the cache blocks are brought into LRSC on an LLC read hit while the LLC still keeps a duplicate copy of block unless the cache block belongs to the IRRA group experiencing zero update memory operation, whereby the HRSC design is chosen for copying the requested block. Accordingly, on a read miss on L1, the associated tag arrays in LRSC and HRSC are simultaneously accessed to look for the missed data block in L2. Even though the parallel search for the missed data block incurs more energy consumption due to enabling two exclusive resources, the overall system performance remains unchanged. These include decoder, tag arrays and wordlines. The energy consumption overhead for these resources has been considered in our simulation results.

To achieve this, the ratio of read and write accesses to LLC cache blocks are monitored through read and write counters. We assign a 6-bit Read Counter (RC) and a 1-bit Write Counter (WC) to each LLC cache block, as justified below. The RC of a cache block in LLC records the read access history to that particular block. This history is utilized to determine if the cache block is a proper candidate to be copied into HRSC. We conducted an extensive exploration to evaluate the preferred value for the read threshold level, $NR_{th}$, within our design. We found that when $NR_{th}$ is small, the ratio of blocks that must be copied into HRSC significantly increases. However, note that the limited capacity of HRSC constrains the number of read-intensive cache blocks that can be maintained. This results in a high ratio of cache blocks to be frequently replaced without providing adequate read services which in turn incurs significant write overhead and undermines the read-friendly property of HRSC. On the other hand, if $NR_{th}$ is too large, then the HRSC utilization significantly decreases because only a few read-intensive cache blocks are selected to be brought into HRSC. Based on our experimental results, $NR_{th}$ equal to \textit{64} maximizes the HRSC utilization while incurring an acceptable cache block replacement rate in HRSC.



The WC determines whether this block has been accessed by a write operation prior to the time that the cache block has reached to $NR_{th}$ or not.  If an LLC line experiences any write operation while resident, it is copied into LRSC upon a read hit on LLC, even if the RC is saturated. Considering 7-bit per line of LLC imposes an acceptable 7-bit/64 Bytes = 1.3\% area overhead.  


Assuming the HRSC is full, one of the lines is evicted based on Least Recently Used (LRU) cache replacement strategy.  Nonetheless, it has been reported in Section \ref{sec:motivation} that 1.55\% of the entire LLC space are IRRA blocks, from which approximately half of them exclusively experience read operation. This means HRSC size is required to be in order of around 0.77\% of the entire LLC size. For example, the HRSC capacity can be less than 800KB to fit all frequently read lines from an LLC with size of 96MB. Thus, we utilized the saved area as a result of replacing SRAM by LRSC in L2 to add facilities for HRSC-based operation. This data arrangement scheme enhances the cache service time because frequently exclusive-read blocks are maintained in the upper-level of cache with the lowest likelihood to be replaced by the` new cache blocks. 



The retention time of STT-RAM cells integrated into LRSC design play a paramount role in determining the energy consumption and performance of the overall L2 design. Thus, we have conducted a study to evaluate the preferred retention time candidates which offer an optimum refresh scheme for LRSC arrangement. 
 \vspace*{-0.1in}

\subsection{Refresh Scheme for LRSC}
The retention time of STT-RAM cell design utilized in LRSC architecture needs to be considered properly to meet the following key design issues: 

1) \textit{data stability during read operation}: The data retention time should be sufficient to retain the stability of data while cache lines are accessed during read operations, otherwise the unstable data is sensed via sense amplifier, which in turn may cause the corrupted data to be provided to the CPU. Even though the sensing resolution and reliability of sense amplifiers employed in STT-RAM cache designs influence the accuracy of the sensed data \cite{zhao2012failure} \cite{jiang2016improving}, other characteristics such as resiliency to process variation \cite{eken2014novel}, performance and power consumption \cite{li2010design}, also play paramount roles for determining the preferred sense amplifier candidate.

2) \textit{competitive performance delivery compared to SRAM-based L2 design}: The employment of high retention STT-RAM cells in the cache design requires a long write pulse for write operation which results in performance degradation. This phenomenon becomes common in write-intensive applications. The retention relaxation of STT-RAM can significantly improve switching performance by reducing the data retention time, which in turn causes reduced write latency.  

3) \textit{cache block accessibility}: To stabilize the data stored in a retention-relaxed cache line, the refresh operation re-writes the cache line which has reached the end of its lifespan. However, if the interval between refreshes is considered to be short, the performance of the system may significantly decrease because the cache block for normal read and write operations is not available during its refresh cycle. In addition, increasing the refresh ratio would undermine the STT-RAM cell's endurance, which is on the order of 10\textsuperscript{12} \cite{jokarsequoia} \cite{TMSCS-wear-leveling}. Thus, the refresh interval should be optimized to reduce the conflict ratio to these cache blocks while incurring insignificant refresh overhead.



\begin{figure}[t] 
\centering
\includegraphics[width=3.5in, keepaspectratio]{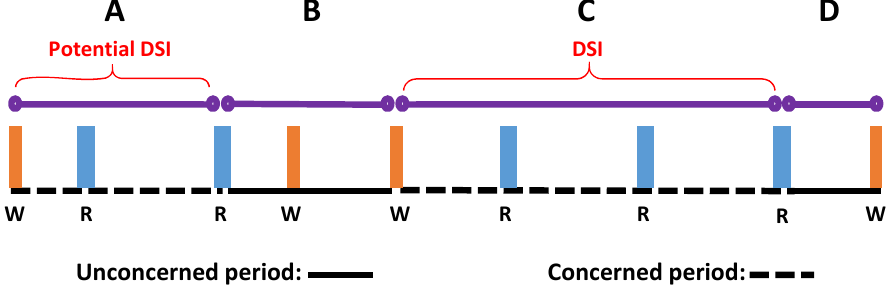}
 \vspace*{-0.1in}
\caption{DSI equals to the sequence C which is largest interval between a write and the final subsequent read operation.}
\label{fig:DSI}
\end{figure}

\begin{figure}[t] 
\centering
\includegraphics[width=3.4in, keepaspectratio]{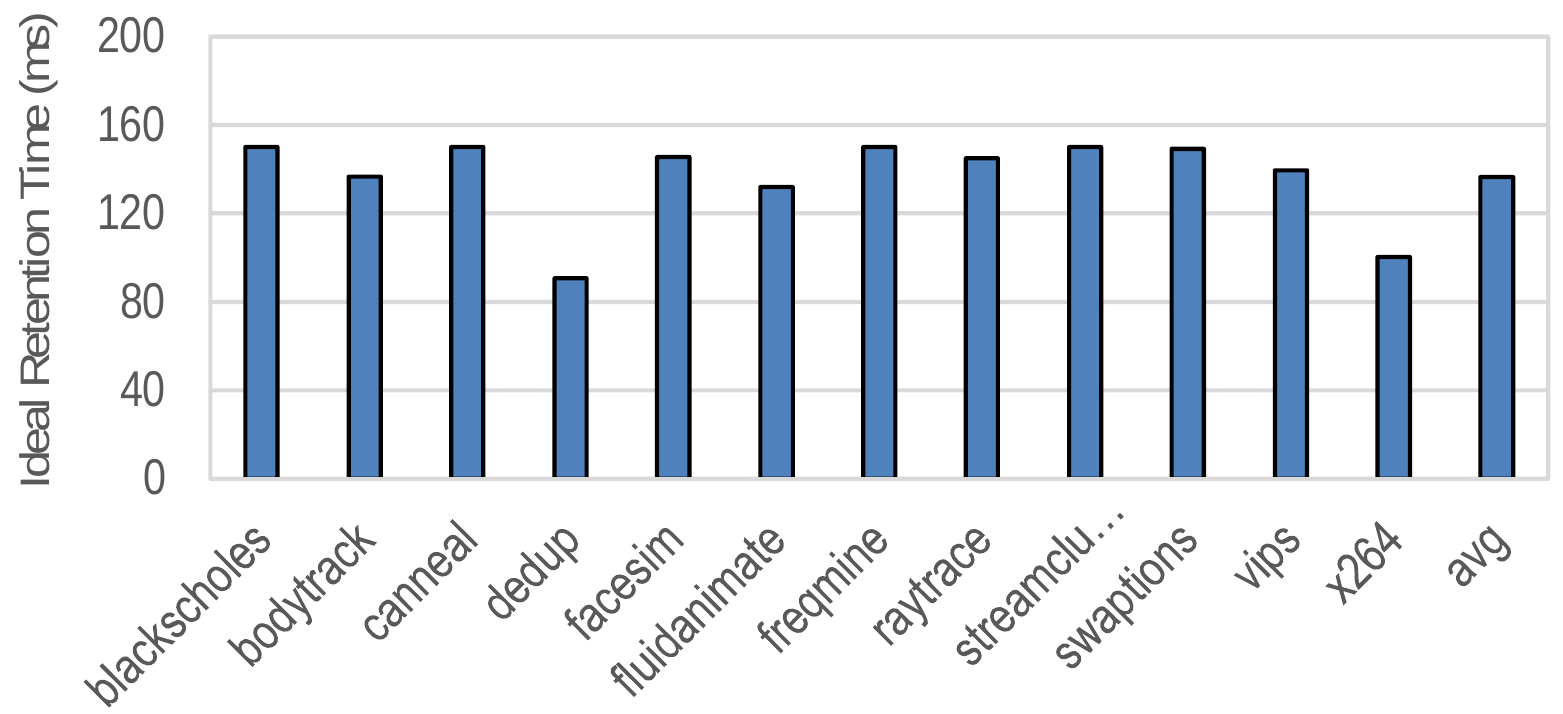}
 \vspace*{-0.1in}
\caption{Ideal data retention time to avoid refresh scheme.}
\label{fig:ideal}
\end{figure}

To find the optimum data retention time which addresses all aforementioned challenges, a new metric called \textit{Data Stability Interval (DSI)} is defined for each cache block. DSI is the maximum interval between a write and the final subsequent read operation before the cache block can be accessed by a write or eviction operation. For instance, Fig. \ref{fig:DSI} shows the entire lifetime of a cache block which has been accessed by multiple read and write memory operations over time. Since the \textit{interval A} begins with a write operation and followed by two read operations, it has the potential to be considered as DSI. The data unstability in the \textit{interval B} does not impact on the processing data in CPU because the data will be over-written during this period. In addition, accessing the cache line by a write operation is similar to refreshing that cache line which guarantees the stability of data up to the end of its lifespan. In \textit{interval C}, the write operation is followed by three read accesses. The \textit{interval C} exhibits longer period compared to \textit{interval A}. Thus, the \textit{interval C} is considered as DSI for this cache block. The ideal data retention time for LRSC design can be defined as follows:

 \begin{equation}
\scriptsize
\label{eqn:ideal_retention_time}
Ideal \quad Retention \quad Time_{LRSC} = Max ( \, DSI_{0}, DSI_{1}, ..., DSI_{n} ) \,
\end{equation}
where \textit{n} is the number of cache blocks. The refresh operation can be completely eliminated for LRSC design with ideal data retention time because the data is stable over the entire concerned period of the cache blocks. Fig. \ref{fig:ideal} shows the ideal data retention time obtained from Eq. \ref{eqn:ideal_retention_time} for both emerging read-intensive and write-intensive workloads selected from PARSEC benchmark suite. Typically, the ideal data retention time for read-intensive workloads is protracted to satisfy the exhaustive read accesses while lengthy read-read sequences are repeatedly taken place. On the other hand, the write-intensive workloads dictate extensive write operations to the cache lines, which intrinsically leads to refreshing accessed cache lines and reduced DSI. Thus, not all cache lines need to be designed ideally for read-intensive and write-intensive workloads because of the non-uniform access behavior to the cache lines. Furthermore, the program runtime behavior is often non-deterministic when running a set of multi-threaded workloads on a multiprocessor architecture which utilizes different branch prediction techniques for performance improvement and avoids unnecessary instruction execution. To clarify this issue, we have conducted an extensive application-driven study to classify DSI distribution over time and to find the optimum data retention time by taking the energy consumption and IPC of different LRSC designs into account.

\begin{figure*}[t] 
\centering
\includegraphics[width=7in, keepaspectratio]{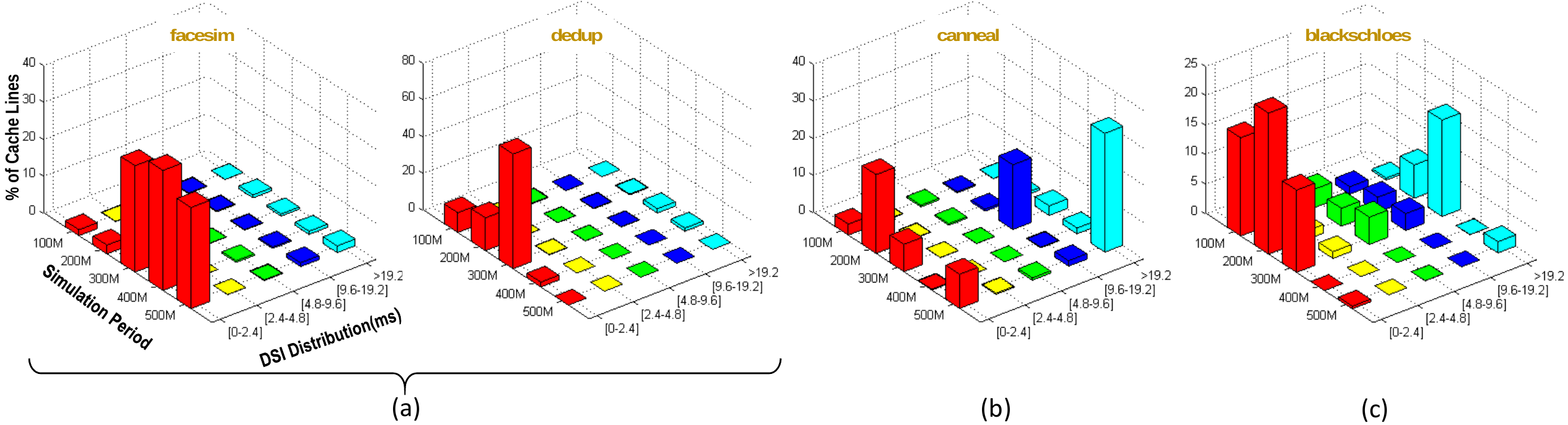}
 \vspace*{-0.2in}
\caption{Three classes of DSI distribution: (a) unimodal, (b) bimodal, and (c) symmetric.}
\label{fig:DSI_dist}
\end{figure*}

Fig. \ref{fig:DSI_dist} illustrates the DSI distribution for four selected workloads. The overall simulation period is divided into five periods and the cache lines are partitioned based on their DSI dispersion. Each bar in the plot depicts what percentage of cache lines with calculated DSI are fall into which simulation period. For instance, less than 5\% of all cache lines in \textit{facesim} need to retain data longer than \textit{19.2ms}, while this quantity goes up to 24.28\% for \textit{blackscholes} workload. Based on the non-uniformity of DSI distribution in different benchmarks, we identify three classes: 

1) \textit{Unimodal DSI Distribution}: A considerable portion of cache lines exhibit short DSI distribution for \textit{facesim} and \textit{dedup} benchmarks, which result in their DSI often falling into the smallest expected period that cache line must preserve data. These benchmarks require relatively narrow window retention time whereby cache lines are accessed by regular write operations. 

2) \textit{Bimodal DSI Distribution}: In this distribution, cache lines are often partitioned into two groups. The first group has short DSI while the DSI of the second group often resides in long-lasting DSI category. The \textit{canneal} demonstrates a bimodal DSI distribution among different cache lines in which around 41\% of cache lines require data retention time less than \textit{2.4ms} while this ratio goes up to 57\% for cache lines with DSI more than \textit{9.6ms}. 

3) \textit{Symmetric DSI Distribution}: The cache lines' DSI are uniformly distributed in each DSI category over simulation periods. The benchmark \textit{blackscholes} has approximately symmetric DSI distribution unlike the behavior of aforementioned benchmarks. 

\begin{figure}[t] 
\centering
\includegraphics[width=3.3in, keepaspectratio]{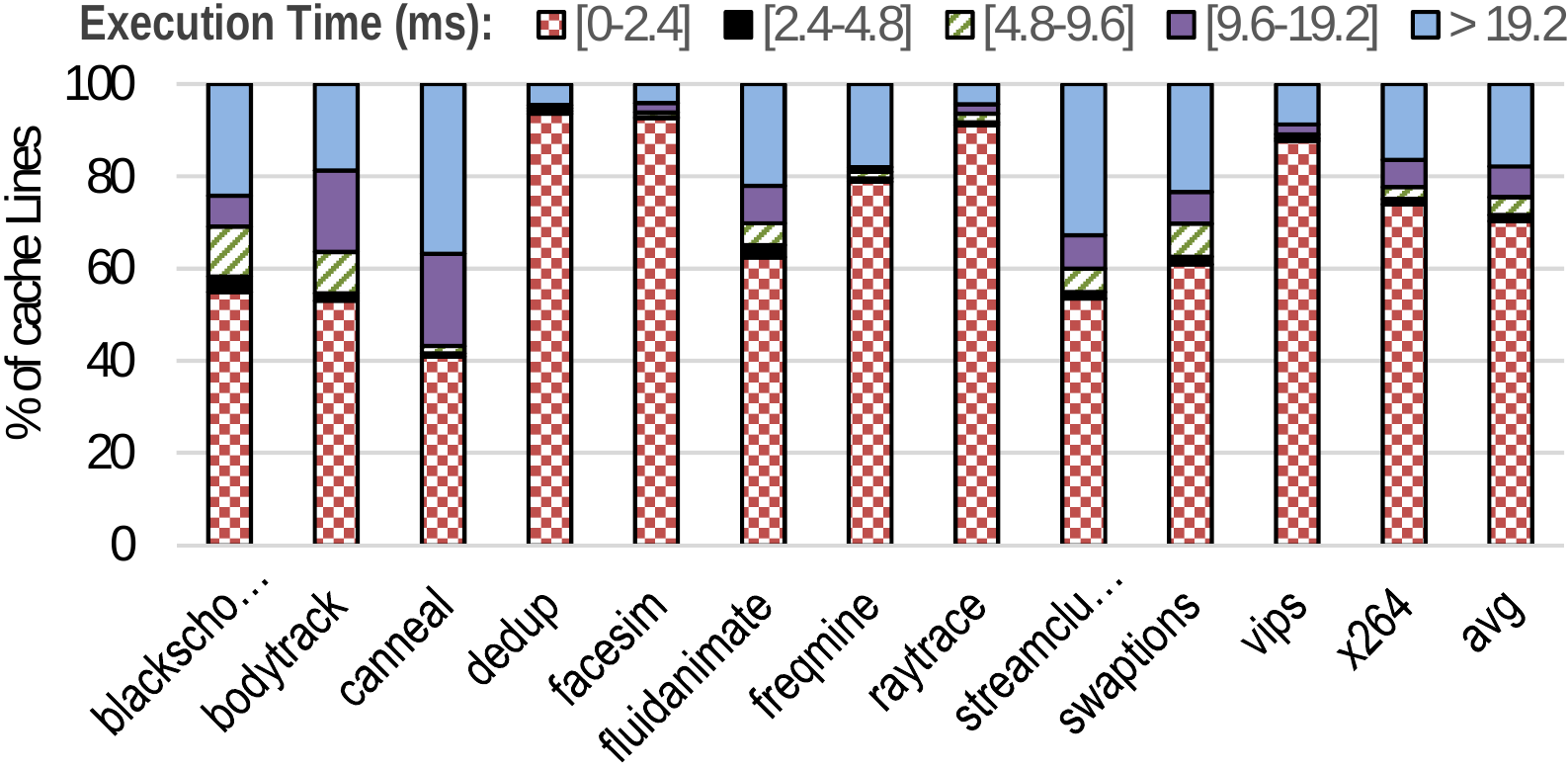}
 \vspace*{-0.2in}
\caption{DSI distribution for all benchmarks.}
\label{fig:DSI_dist_all}
\end{figure}

Fig. \ref{fig:DSI_dist_all} shows the total cache lines distribution with different DSI, where the majority of cache lines (on average 70.3\%) require data retention time less than \textit{2.4ms}. On the other hand, around 18\% of cache lines need long-lasting retention time more than \textit{19.2ms} to meet data stability purpose during read operation. 

To find the sufficient data retention time that accommodates provision for the above three classes while miniaturizing conflict with normal memory accesses and delivering high performance, we selected three different STT-RAM cell's retention time for comparison in terms of incurred energy consumption and offered performance as listed in Table \ref{tab:LRSC_design}. \textit{Design 1} complies with the demands of both \textit{bimodal} and \textit{symmetric DSI distributions} to provide data retention time as high as ideal approach for cache lines with DSI more than \textit{19.2ms}. \textit{Design 3} offers a reduced retention time in favor of workloads with \textit{unimodal DSI distribution} to promote write performance. \textit{Design 2} has been considered as an intermediary to satisfy both groups of cache lines claiming either long or short DSI while also targeting cache lines having middle retention time. The reduced data$'$s lifespan stored in \textit{Design 2} and \textit{Design 3} require a refresh mechanism to prevent data loss. Therefore, the proposed refresh scheme in \cite{sun-micro-2011} is considered to sequentially refresh all cache blocks. We have included the energy contributions from peripheral circuits and energy consumption due to refresh mechanism in our simulation results. The detailed scheme for adjusting the retention time of the STT-RAM cell is elaborated in Section \ref{sec:relax_LRSC}.

\begin{table}
\renewcommand{\arraystretch}{1.3}
\scriptsize
\caption{STT-RAM cell retention time configurations for LRSC design.}
 \vspace*{-0.1in}
\label{tab:LRSC_design}
\centering
\begin{tabular}{|c|c|c|c|}
    \hline
    Configuration &  Design 1 & Design 2 & Design 3 \\
      \hline
    Retention Time  &   140ms	& 10ms & 1ms \\
    \hline
	Write Latency  @3GHz   & 12 cycles & 7 cycles & 6 cycles \\
	\hline
\end{tabular}
\end{table}

\begin{figure*}[t] 
\centering
\includegraphics[width=6.5in, keepaspectratio]{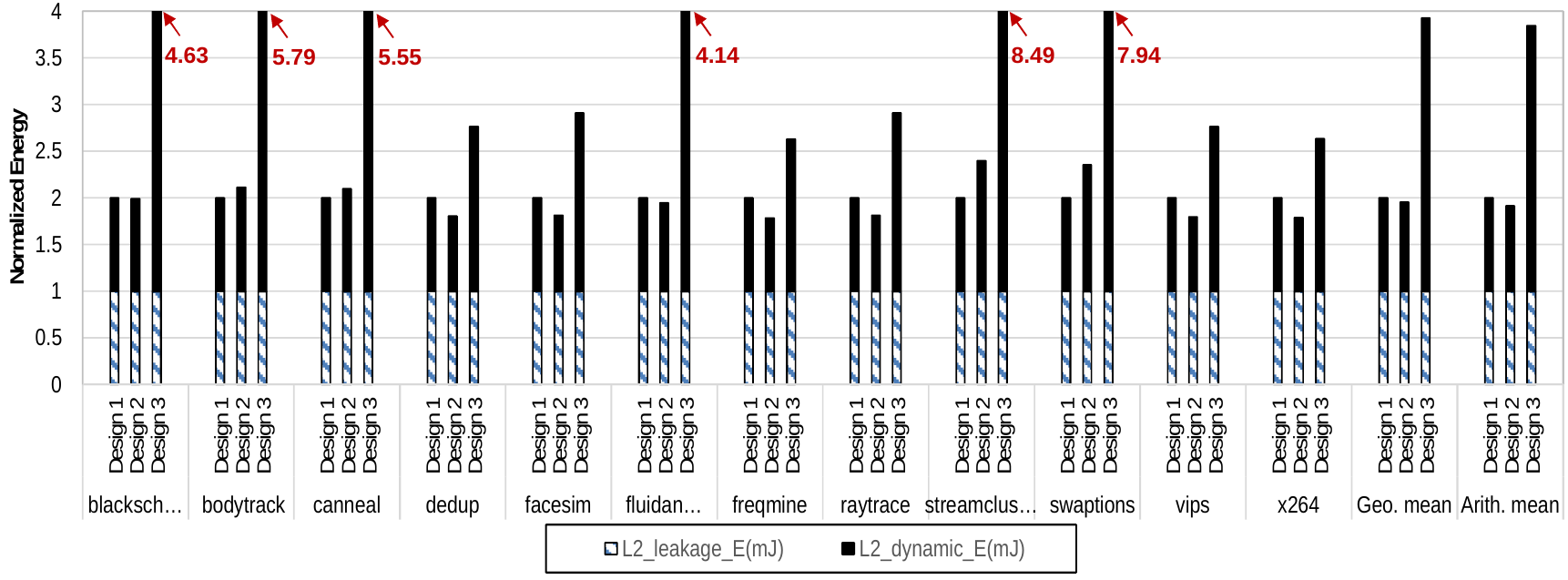}
 \vspace*{-0.1in}
\caption{Energy breakdown comparison among three LRSC configurations normalized to Design 1.}
\label{fig:energy_d1_d2_d3}
\end{figure*}

\begin{figure}[t] 
\centering
\includegraphics[width=3.5in, keepaspectratio]{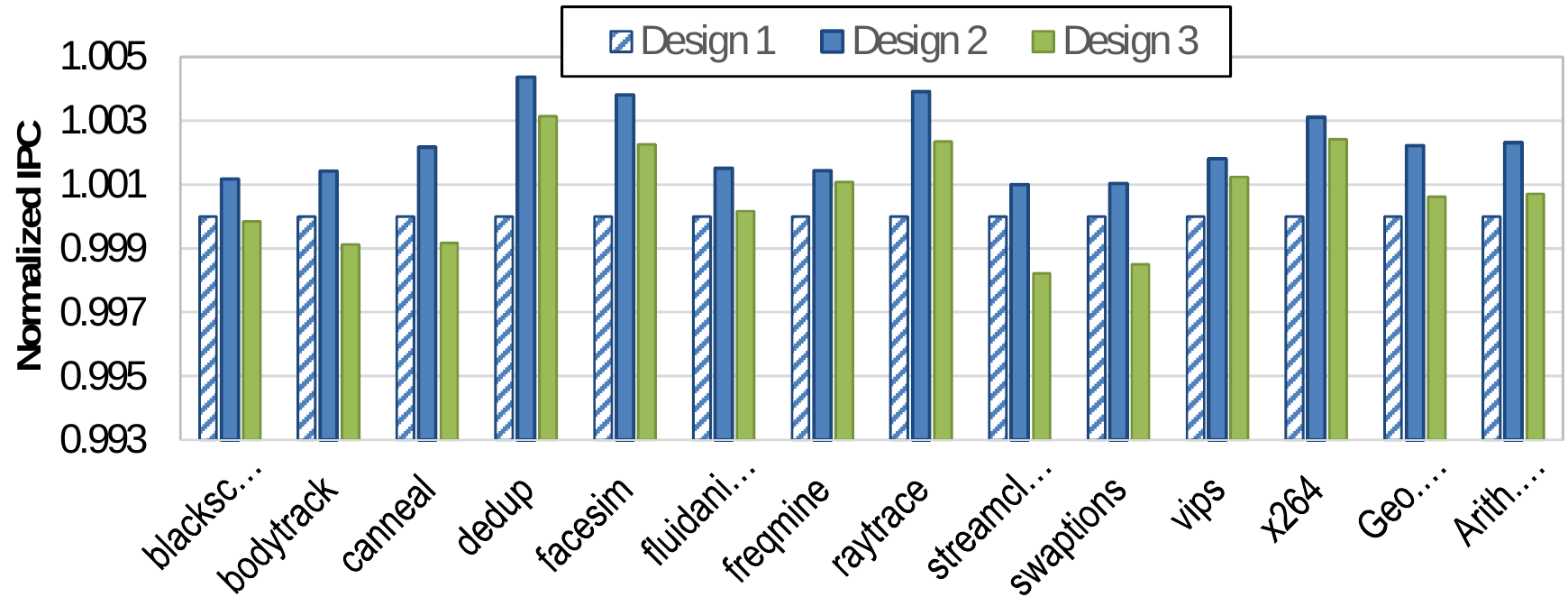}
 \vspace*{-0.1in}
\caption{IPC comparison among three LRSC configurations normalized to Design 1.}
\label{fig:IPC_d1_d2_d3}
\end{figure}

The energy consumption and IPC comparison for the above three designs are shown in Fig. \ref{fig:energy_d1_d2_d3} and Fig. \ref{fig:IPC_d1_d2_d3}, respectively. The relatively high retention time of \textit{Design 1} comes at the expense of high write current and slow write speed while completely eliminating the energy dissipation and memory access conflict induced by periodic refresh operation. \textit{Design 2} leverages the small write current realized by lowering retention time to diminish dynamic energy consumption and to improve write performance. Although it is expected to observe decent improvement in both criteria compared to previous design, the results show slight gain. The main reason is that the extra write operations associated with periodically refreshing cache lines incur additional energy dissipation while the conflict between refresh operations and normal memory accesses is trivial.     
The augmented number of refreshes for \textit{Design 3} would undermine the benefits of volatile STT-RAM with \textit{1ms} retention time whereby the conflict ratio among refresh operations and normal read/write accesses significantly increases besides advancing the write speed. 
 Thus, since the \textit{Design 2} offers lowest energy consumption and highest IPC among three designs, we selected it as the basis for our LRSC design. \textit{Design 2} benefits from reduced retention time to improve energy consumption and delivers competitive write performance, while the negative impact of its refresh operations is amortized. \textit{Design 2} reduces energy consumption by 57.09\% on average compared to \textit{Design 3} and enhances the overall IPC slightly in comparison with \textit{Design 1}. These experimental results confirm the conducted study in \cite{jog-dac-2012}.

\subsection{Retention  Relaxation in LRSC Design}\label{sec:relax_LRSC}
STT-RAM write performance improvement relates to the optimization of retention time \cite{jog-dac-2012}\cite{sun-micro-2011}. STT-RAM is usually considered as non-volatile technology, while its non-volatile characteristics can be relaxed to obtain better write performance. The retention time of STT-RAM, which is the period that STT-RAM can retain data until a bit-flip occurs, can be modeled as \cite{chang-hpca-2013}:  
 \vspace*{-0.1in}
\begin{equation}
\label{eq:retain_data}
t = t_{1} \times e^\Delta
\end{equation}
where $t$ is the retention time, $t_{1}$ is is fitting constant, and $\Delta$ is thermal barrier that determines the stability of STT-RAM. The retention time of STT-RAM can be reduced exponentially by using the thermal barrier reduction whereby $\Delta$ can be characterized using \cite{kang-aspdac-2014}:
 \vspace*{-0.1in}
\begin{equation}
\label{eq:thermal_barrier}
\Delta \approx {M_{s} H_{k} V \over T}
\end{equation}
where $M_{s}$ is the saturation magnetization, $H_{k}$ is the in-plane anisotropy field, $V$ is the volume of free layer, and $T$ is the absolute temperature in Kelvin.

It has been demonstrated in \cite{kang-aspdac-2014} that there are two STT-RAM switching modes: 1) \textit{Thermal  Activation (TA) region}, where $I_{write}$ is less than critical write current ($I_{c}$) and 2) \textit{Precessional Switching (PS) region}, where $I_{write}$ is greater than $I_{c}$. In PS mode, the write pulse width reduction incurs the rapid increase of write current. Accordingly, some particular write pulse width can deliver optimal STT-RAM write energy. The focus of this paper is on overall write latency and energy optimization in L2 structure design. Thus, herein we adjust $\Delta$ in Eq. \ref{eq:switch_mode} to 1) calibrate the STT-RAM write speed such that it is comparable with SRAM, 2) find the optimal write current for shorter write pulse width which can still provide required write energy needed for switching in the PS region. The switching duration can be calculated as follows \cite{worledge2011spin}:  
\begin{equation}
\label{eq:switch_mode}
{1 \over \tau_{1}} = \big [ {2 \over (C+ln(\pi^2 \Delta))} \big]  {\mu_{B}P \over em(1+P^2)} (I_{write}-I_{c}) 
\end{equation}
where $\tau_{1}$ is the switching mean duration, $C=0.577$  is  
Euler$'$s constant, $\mu_{B}$ is  the  Bohr  magneton constant, $P$ is the tunneling spin polarization of the ferromagnetic layers, $e$ is is the magnitude  of  the  electron  charge,  $m$ is the free layer magnetic moment, and $I_{c}$ is critical write current computed as below:
 \vspace*{-0.1in}
\begin{equation}
\label{eq:critical_current}
{I_{c}} = 2 \alpha {\gamma e \over \mu_{B}g} E
\end{equation}
where $\alpha$ is Gilbert damping coefficient, $\gamma$  is the Gyro-magnetic constant, $g$ is the spin polarization efficiency factor and $E$ is the barrier energy \cite{kang-aspdac-2014}.

\begin{figure}[t] 
\centering
\includegraphics[width=3.3in, keepaspectratio]{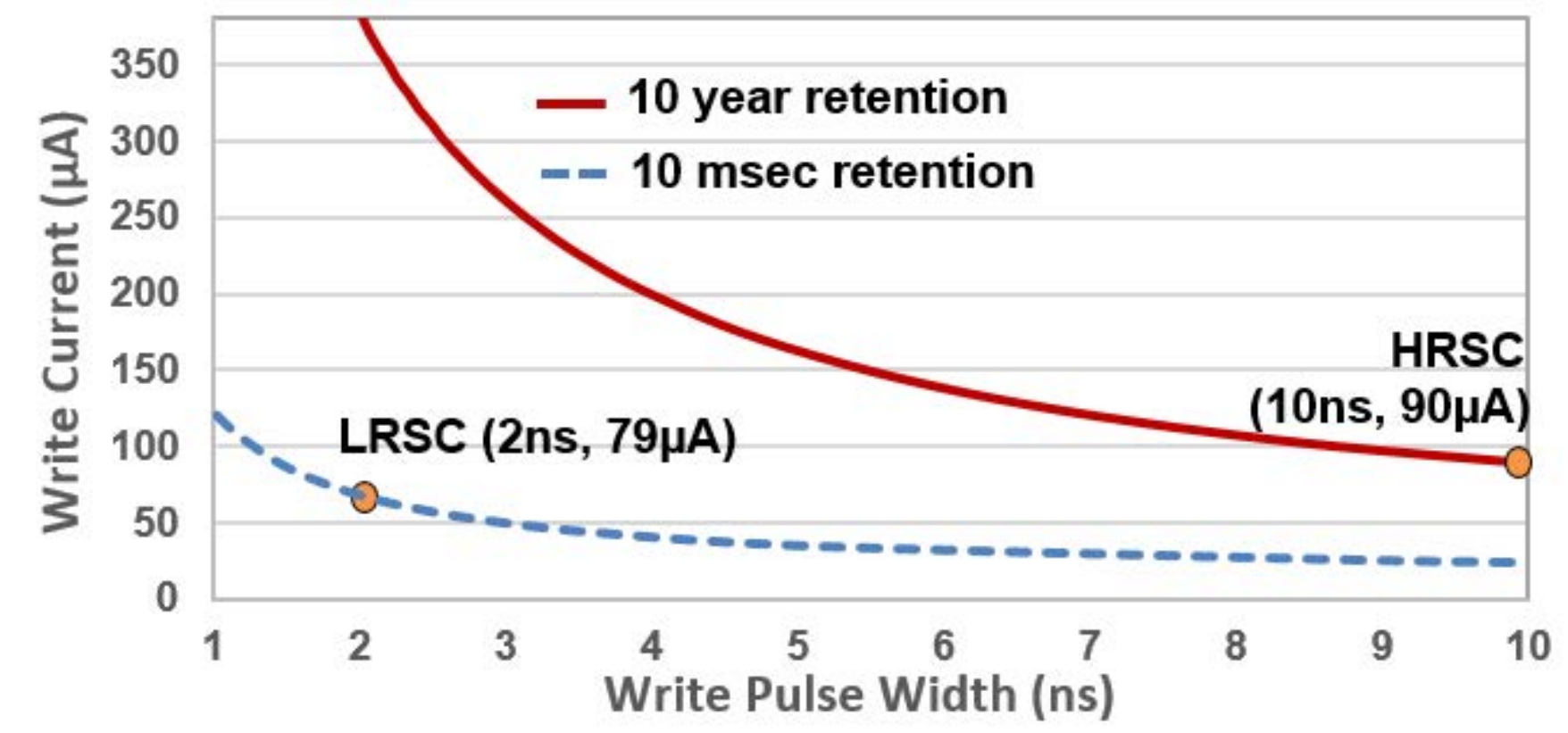}
 \vspace*{-0.2in}
\caption{Write current vs. write pulse width for LRSC and HRSC.}
\label{fig:optimum-write}
\end{figure}

RRAP utilizes two distinct memory arrays to provide categories of retention times matching the required memory reference characteristics as depicted in Fig.~\ref{fig:optimum-write}.  In our study, STT-RAM with a ten year retention time and high $\Delta$ is considered as the baseline. The volatile STT-RAM with lower $\Delta$  offers retention time of \textit{10ms}.
As shown in Fig. \ref{fig:optimum-write}, two operating points HRSC (10ns, 90$\mu A$) and LRSC (2ns, 79$\mu A$) are selected. Based on the RRAP methodology, LRSC requires to be operated with low retention time while high retention time STT-RAM is needed for HRSC design. In order to reduce thermal barrier to achieve a low retention STT-RAM cell, we utilized the previous methodology proposed in previous works \cite{jog-dac-2012} \cite{sun-micro-2011} in which the planar area and thickness of STT-RAM is reduced resulting in exponentially reducing the retention time and required write current for switching. The write current, write pulse width, and the cell size associated to each point are integrated into NVSim \cite{dong-tcad-2012} to obtain energy consumption and latency factors. The detailed device characterization of L2 cache structure are listed in Table \ref{tab:L2_conf}. 
 \vspace*{-0.1in}


\section{Experimental Evaluation}\label{sec:experiment}

\subsection{Simulator Configuration}
We evaluate our design using MARSSx86 \cite{mars}  with PARSEC 2.1 applications. We model a Chip Multi-Processor (CMP) with eight single-threaded x86 cores. Each core consists of private L1, LHRSC L2 (LRSC and HRSC) and the LLC shared among all the cores. The detail of our model can be found in Table \ref{tab:mem_sub}. Twelve applications from the PARSEC suite are selected and executed 500 million instructions starting at the Region Of Interest (ROI) after warming up the cache for 5 million instructions. The {\tt simsmall} input sets are used for all PARSEC applications. The latency and energy usage associated with read and write operations for SRAM and STT-RAM cache accesses are provided by NVSim while DESTINY \cite{Poremba-date-2015} is used to model eDRAM model. This is because NVSim model of eDRAM is incomplete and has not been validated. In addition, the energy contributions from peripheral circuits are also included in our simulation.

\begin{table}[b]
\renewcommand{\arraystretch}{1.3}
\scriptsize
\caption{Detailed characteristics of private L2 cache bank configuration
(32nm, temperature=350K)}
 \vspace*{-0.1in}
\label{tab:L2_conf}
\centering
\begin{tabular}{|c|c|c|c|c|c|c|}
    \hline
    L2 Cache &  Area & RL & WL & RE & WE & LP\\
    Technology&$(mm^2)$& $(ns)$&$(ns)$&$(nJ)$&$(nJ)$&$(mW)$  \\
      \hline
    512KB   &   1.410	& 1.277 & 1.277 & 0.293 & 0.293 & 1753.444\\
     SRAM&&&&&&\\
    \hline
	1MB eDRAM     & 0.745 & 1.072 & 1.022 & 0.289  & 0.424 & 337.329\\
	\hline
    1MB STT-RAM     & 0.526 & 1.340 & 10.218 & 0.280  & 0.654 & 212.022\\
	\hline
    512KB LRSC     & 0.243 & 1.260 & 2.153 & 0.233  & 0.269 & 104.797\\
    \hline
	512KB HRSC     & 0.357 & 1.261 & 10.153 & 0.233  & 0.601 & 114.915\\
	\hline
\end{tabular}

RL: Read Latency, WL: Write Latency, RE: Read Energy,

WE: Write Energy, LP: Leakage Power
\end{table}
 
 
\subsection{Energy Usage Comparison}
The detailed characteristics of considered technologies to be integrated into L2 design are listed in Table \ref{tab:L2_conf}. In order to evaluate the energy benefit of RRAP, we compare the energy breakdown of RRAP with 512KB SRAM-based L2, 1MB eDRAM-based L2 and regular 1MB STT-RAM with long retention time. Note here RRAP consists of 512KB LRSC (Design 2 in Section 4.2) and 512KB HRSC.  


 
\begin{figure*}[t] 
\centering
\includegraphics[width=5.5in, keepaspectratio]{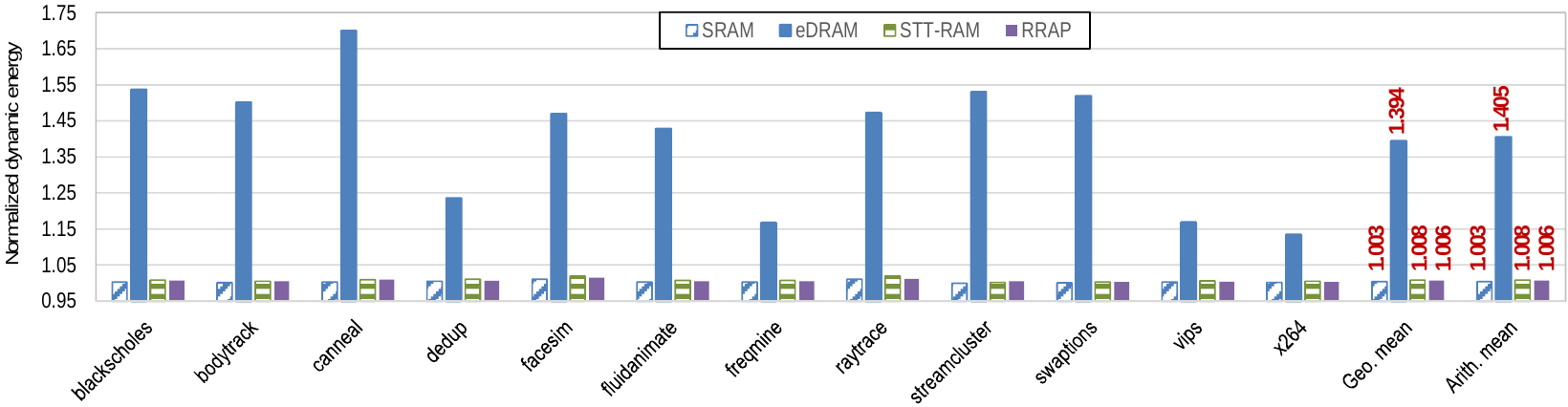}
\vspace*{-0.1in}
\caption{L2 dynamic energy breakdown for SRAM, eDRAM, regular STT-RAM and RRAP.} 
\label{fig:L2_power_dynamic}
\end{figure*}
\begin{figure*}[t] 
\centering
\includegraphics[width=5.5in, keepaspectratio]{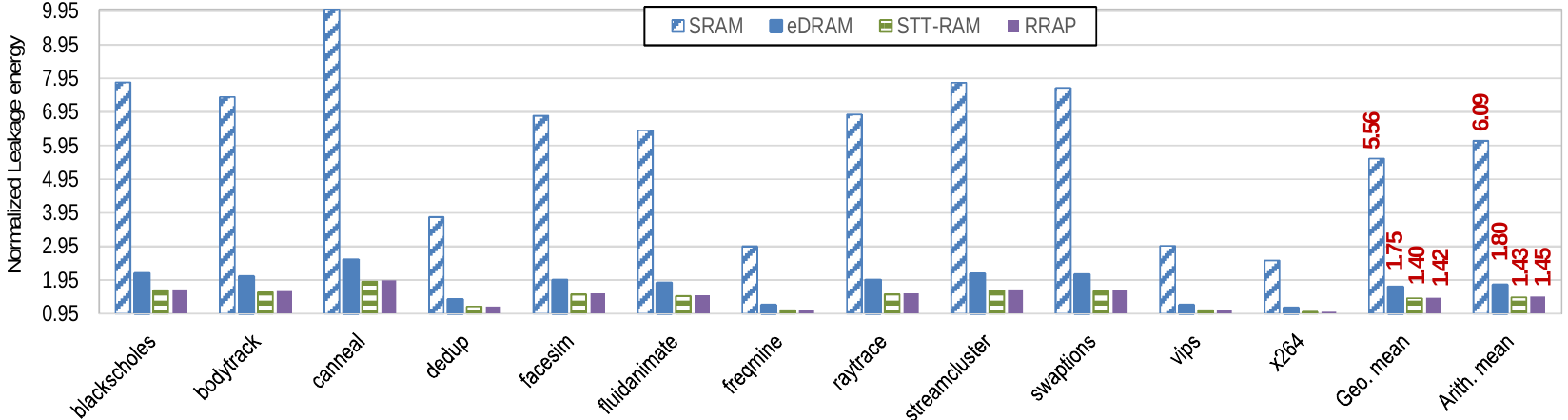}
\vspace*{-0.1in}
\caption{L2 leakage energy breakdown for SRAM, eDRAM regular STT-RAM and RRAP.} 
\label{fig:L2_power_leakage}
\end{figure*}

 The write operations in HRSC and regular STT-RAM impose long write latency because high energy write pulse is required to retain the data for a longer time compared to two other technologies. On the other hand, the data$'$s lifespan stored in LRSC is \textit{10ms}, which makes it necessary to employ a refresh mechanism to prevent data loss. Therefore, a low-overhead refreshing scheme introduced by \cite{sun-micro-2011} is deployed here in which all cache blocks are refreshed sequentially. Consequently, as shown in Fig. \ref{fig:L2_power_dynamic}, the consumed dynamic energy of RRAP approach is slightly higher than the dynamic energy consumed by SRAM-based L2, but it is still significantly less than dynamic energy consumption of eDRAM which demands exhaustive refresh operation every $40us$ \cite{chang-hpca-2013} \cite{khoshavi_isqed_2016}. Note that the high energy required for write operation in HRSC has negligible effect on the overall dynamic energy consumption of RRAP because the cache blocks are brought into HRSC once and are accessed by read operations for the rest of execution time. Moreover, the high write energy and large memory cell size in regular STT-RAM incur higher dynamic energy consumption in comparison with RRAP. With RRAP, 97.6\% of dynamic energy is reduced compared to eDRAM-based L2 design on average.
Fig. \ref{fig:L2_power_leakage} compares the leakage energy consumption for aforementioned designs. The regular STT-RAM and RRAP offer the least leakage energy compared to SRAM and eDRAM-based L2 because STT-RAM is highly persistent against leakage. On the other hand, SRAM consumes relatively higher leakage energy than other designs. 


\vspace*{-0.1in}

\begin{table}
\renewcommand{\arraystretch}{1.3}
\scriptsize
\caption{Memory subsystem}
 \vspace*{-0.1in}
\label{tab:mem_sub}
\centering
\begin{tabular}{|c|c|}
    \hline
    Processor  &  3GHz processor Fetch/Exec/Commit width 4\\
    \hline
    Private L1-I/D  &SRAM, 32 KB, 8-way set assoc., WB cache \\
    \hline
    Private L2 Conf.  & 8-way set assoc., WB cache\\
    \hline
	Shared L3  &eDRAM, 96 MB, 16-way set assoc., \\
	& 16 bank, WB cache\\
	\hline
	Main memory & 8 GB, 1 channel, 4 ranks/channel, 8 bank/rank\\
	\hline
\end{tabular}
 \vspace*{-0.1in}
\end{table}

\subsection{Read Miss Ratio Comparison}

\begin{figure}[ht] 
\centering
\includegraphics[width=3.5in, keepaspectratio]{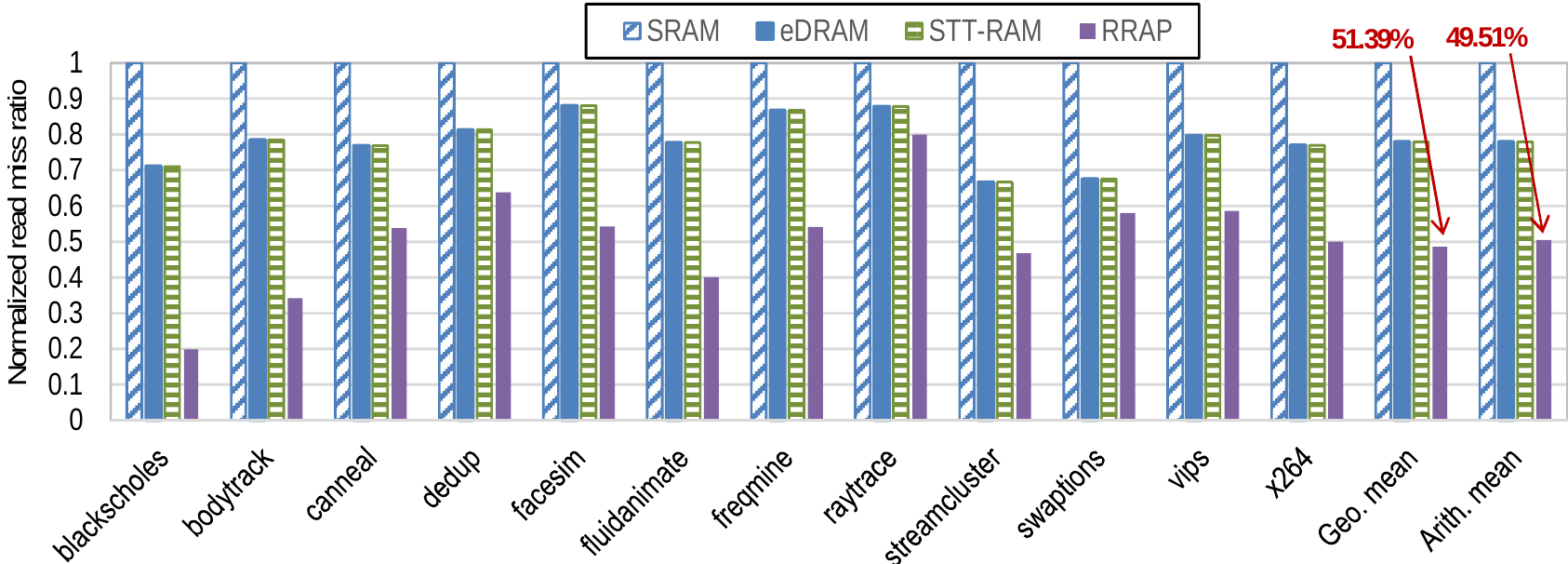}
\vspace*{-0.3in}
\caption{Comparison of L2 read miss ratio normalized to SRAM.}
\label{fig:read_miss}
\end{figure}

In light of RRAP, the read miss ratio of L2 is significantly decreased as illustrated in Fig. \ref{fig:read_miss}. RRAP enhances the read miss ratio of the SRAM-based L2 design by 51.39\% on average (geometric mean). The experimental results show that the read-intensive workloads, such as \textit{blackscholes}, \textit{bodytrack} and \textit{streamcluster}, leverage the full potential of RRAP to further diminish the read miss ratio compared to eDRAM-based and regular STT-RAM L2 designs. One of the main reasons that RRAP can achieve the reduced miss ratio is the provided extra cache capacity due to deploying STT-RAM instead of SRAM. The other reason is the capability of RRAP approach for bringing frequently exclusive read accessed blocks from LLC to HRSC in upper-level of cache while guaranteeing the cache blocks reside for a large window of execution time.

\subsection{Read Service Time Comparison}
\begin{figure}[ht] 
\centering
\includegraphics[width=3.5in, keepaspectratio]{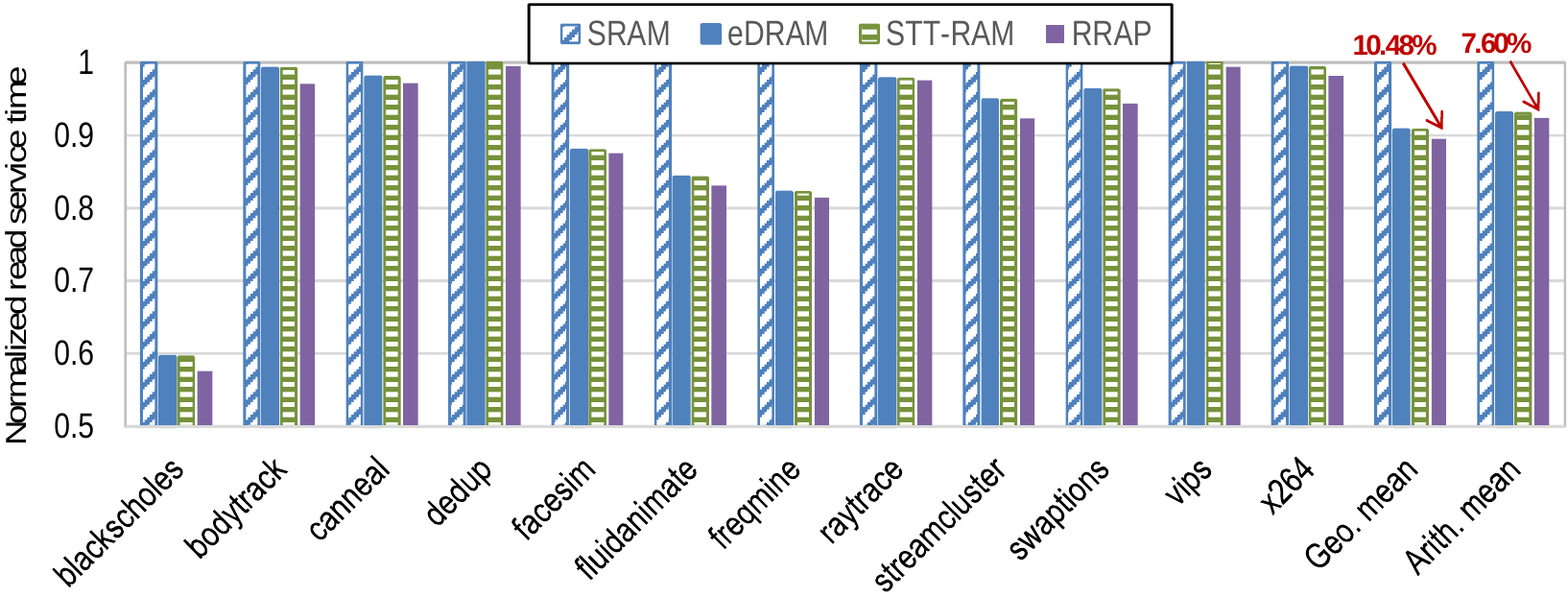}
\vspace*{-0.3in}
\caption{Comparison of read service time for the entire memory hierarchy normalized to SRAM.}
\label{fig:read_service}
\end{figure}

Herein, the elapsed time to service a read request by the processor is referred to Read Service Time (RST). To estimate the RST for the entire system, we have considered the amount of read hits and misses in all three levels of cache while also taking the trip latency for transferring a cache block from lower-level to upper-level of cache into account. The SRAM-based L2 design offers the longest RST compared to other designs as shown in Fig. \ref{fig:read_service}. The reason is primarily because of its small L2 capacity while using area-inefficient SRAM cells for integrating into L2 structure.

On the other hand, the designs which utilize eDRAM and regular STT-RAM in L2 offer nearly comparable RST compared to RRAP technique. The main reason is that the service time to bring a data block from main memory to LLC contributes the most to the overall RST. This negatively impacts on the efficiency of RRAP which decreases the read miss ratio in L2. Another reason is that all three techniques leverage double-sized cache capacity for L2 in comparison with SRAM-based design to satisfy the high demand for maintaining read-intensive cache blocks. Nonetheless, compared to eDRAM and regular STT-RAM designs, RRAP provides 60.4\% and 1.3\% on average less RST for the read requests from L2 and the total read requests, respectively. 



\vspace*{-0.1in}
\subsection{Performance Comparison}
\begin{figure}[ht] 
\centering
\includegraphics[width=3.5in, keepaspectratio]{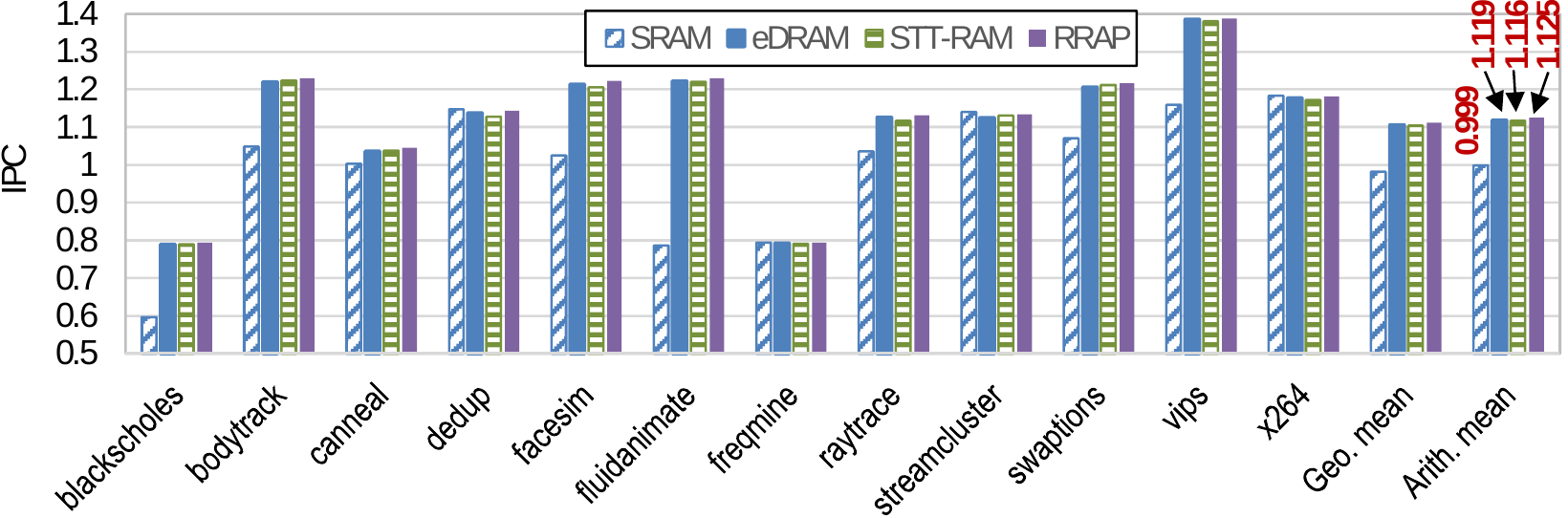}
\vspace*{-0.3in}
\caption{Comparison of RRAP's performance with other technologies.}
\label{fig:speedup}
\end{figure}

Besides the read miss ratio and RST comparison, we also consider IPC to compare the overall performance. Fig. \ref{fig:speedup} illustrates the IPC of the systems where eDRAM-based L2, regular STT-RAM and RRAP exhibit greater performance compared to SRAM-based L2 design. RRAP achieves better performance due to the reduced miss ratio in comparison with eDRAM-based and regular STT-RAM designs. The results show that eDRAM-based L2 and regular STT-RAM designs offer nearly identical performance. The main reason for performance degradation in regular STT-RAM is its high write latency compared to the same operations' latency in eDRAM-based L2 and LRSC designs. The cache blocks in HRSC are accessed by write operation only one time while the remained memory accesses are read operations. Thus, the high write latency in HRSC has negligible effect on performance degradation of RRAP. In addition, RRAP leverages an efficient retention-relaxed STT-RAM as LRSC to amortize the incurred high latency associated with the write operation in STT-RAM cells. The conflict between refresh operations and normal read/write accesses are the major source of performance degradation in eDRAM-based L2 design.



The results show that RRAP improves the system performance by 11.7\% on average (geometric mean) compared to the SRAM-based L2, which is mostly due to employing larger L2 cache capacity for managing cache requests and bringing LLC frequently read blocks to high retention L2 design.

\vspace*{-0.1in}

\section{Related Work}
The hybrid cache design approaches have received significant attention over past years. These techniques leverage the benefits of utilized technologies to maximize the performance and minimize the energy consumption. In \cite{valero2015design}, both SRAM and eDRAM technologies are employed in the second-level cache to offer area savings and reduced energy consumption while the performance is increased by 5.9 percent on average. The proposed LLC design leverages the fast SRAM device to store the most likely referenced blocks in the future while the high-dense and low-leakage eDRAM is utilized to reduce the overall energy consumption \cite{valero2015design}. In other words, the larger SRAM banks can provide the higher performance while the deployment of larger area-efficient eDRAM requires lower energy for operation. Thus, the ratio of utilizing these technologies in the LLC has been optimized to provide the sufficient performance and reduce energy consumption. However, this design does not explicitly consider the critical load behavior.
 
The proposed work in \cite{Imani2016} leverages the asymmetric write power associated with storing 'ones' and 'zeros' to place data blocks having majority 'zero' data into STT-RAM while the remained data blocks are stored in the SRAM. To deal with the change of majority due to regular update operations, a two-bit counter has been considered for each cache line to track the status of the majority. To avoid write overhead associated with the unnecessary migration, the migration policy swap the data blocks according to the pace of majority-data changes. This design can limit the performance of multicore designs to the efficiency of the module which determines the majority of the data. Namely, each data block should be given to the dispatcher module before placement in the LLC which incurs significant overhead for memory-intensive applications and impose significant pressure on the placement policy.


The proposed hybrid LLC design proposed in [3] utilizes an intelligent adaptive data block placement by taking each cache line future access pattern into consideration. 
 The write accesses are categorized into three main classes: prefetch-write, core-write and demand-write. Around 26\% of all prefetch blocks are not accessed by the core after initial prefetch \cite{wang2014adaptive}. Furthermore, the data blocks moved to the LLC due to cache burst phenomenon are not accessed again until the eviction occurs \cite{wang2014adaptive}. The data blocks with the aforementioned characteristics are considered to be placed in the SRAM. 
 On the other hand, the data blocks which experience long interval between consecutive reads are placed into the STT-RAM to benefit from the low leakage power cost of long-residency offered by non-volatile devices \cite{wang2014adaptive}. This design may incur prediction design overhead which incurs extra energy consumption for tracking the access pattern to each cache line and making the decision for placement.

In \cite{sun-micro-2011}, Low-Retention (LR) and High-Retention (HR) STT-RAM arrays are utilized simultaneously to balance the performance versus the energy consumption. To accomplish this, the retention time of the STT-RAM is relaxed for LR architecture to improve the write operation speed. On the other hand, the HR cache offers the long-term residency of data block while incurring very small leakage power. Regarding to the features that each cache design offers, the proposed method manages the write-intensive cache blocks to be placed into the LR cache in favor of performance, while the read-intensive cache blocks are kept in the HR arrays to meet the required power budget limits. Furthermore, the migration policy is devised to transfer the data blocks to the proper cache design by continuously monitoring the access pattern to each cache line. Since the lifespan of the cache lines in the LR arrays is limited to the retention time which is deliberated at the design time, a refresh mechanism is designed to periodically refresh cache lines for preventing data loss. RRAP utilizes the same refresh approach to maintain the stability of data in LRSC design.

\begin{figure}[ht] 
\centering
\includegraphics[width=3.5in, keepaspectratio]{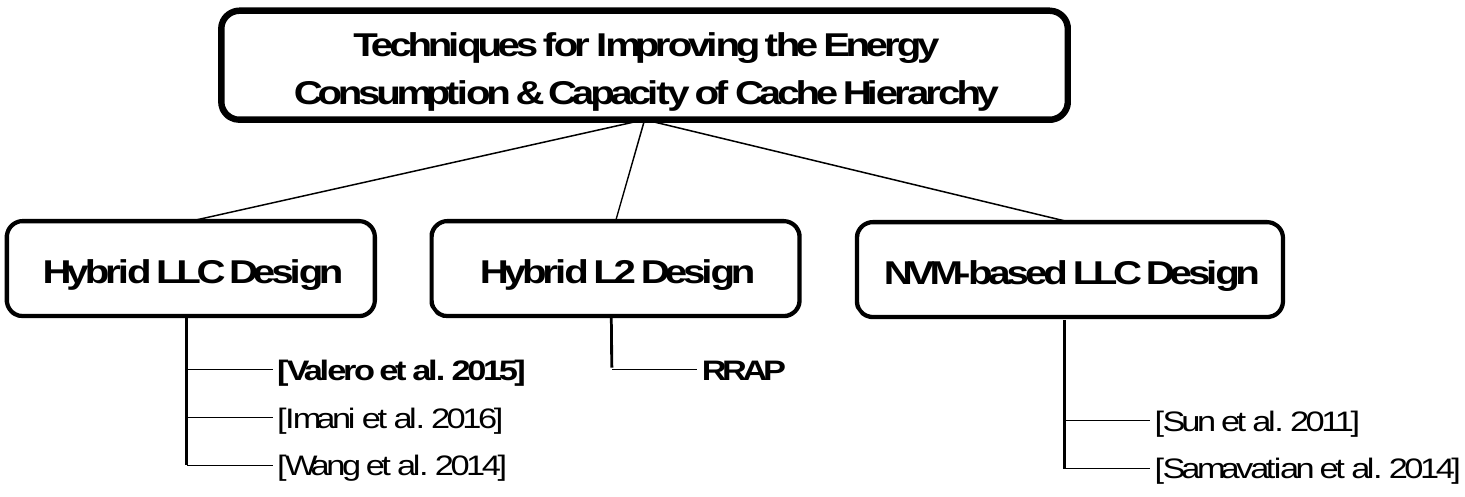}
\vspace*{-0.1in}
\caption{Taxonomy of techniques utilizing the emergent technologies for improving the performance and energy consumption. Approaches using eDRAM highlighted in bold face font.}
\label{fig:taxonomy}
\end{figure}

In \cite{samavatian2014efficient}, the similar idea is utilized in the LLC design for GPU. The small-sized LR cache is employed to keep the write-intensive data blocks by considering the fact that the re-write interval time of blocks is typically lower than $100\mu s$.  Again, the large HR arrays is considered to maintain the less-frequently written data blocks \cite{samavatian2014efficient}. The proposed technique uses a write threshold on HR to determine whether to keep the data in HR or move it to the LR.  To achieve this goal, the correlation between the threshold value and the performance has been accurately analyzed, and the optimum value has been selected as the threshold value. In addition, different degrees of associativity are considered for LR and HR designs to maximize the write utilization in LR and to improve the data migration policy. The taxonomy of the discussed techniques are presented in Fig. \ref{fig:taxonomy}.

\vspace*{-0.1in}

%
%

\section{Conclusions}
In conclusion, RRAP retains the on-chip cache utilization close to the requesting cores for data locality maximization and lower memory access latency while taking into account the energy consumption and performance speed-up. HRSC design leverages the non-volatility of STT-RAM to store IRRA cache blocks, which in turn provides long-term residency of data without demanding extra energy to maintain the data stability. For LRSC design, we conducted an extensive exploration to find the preferred configuration in terms of energy consumption and performance. Accordingly, the write latency of LRSC is relaxed to maintain the cache utilization as high as SRAM attains. Our experimental results show that RRAP can reduce the overall energy consumption and read miss ratio significantly, and read service time slightly within a comparable footprint to SRAM-based L2 designs. 


\section*{Acknowledgment}
This work is supported in part by the US National Science Foundation Grant CCF-1527249, CCF-1337244 and National Science Foundation Early Career Award 0953946. The authors would also like to thank Brianna Thomason for valuable review and suggestions.

\ifCLASSOPTIONcompsoc
\else
\fi


\ifCLASSOPTIONcaptionsoff
  \newpage
\fi

\bibliographystyle{IEEEtran}
\bibliography{IEEEabrv,bibfile}

\begin{thebibliography}{10}
\providecommand{\url}[1]{#1}
\csname url@samestyle\endcsname
\providecommand{\newblock}{\relax}
\providecommand{\bibinfo}[2]{#2}
\providecommand{\BIBentrySTDinterwordspacing}{\spaceskip=0pt\relax}
\providecommand{\BIBentryALTinterwordstretchfactor}{4}
\providecommand{\BIBentryALTinterwordspacing}{\spaceskip=\fontdimen2\font plus
\BIBentryALTinterwordstretchfactor\fontdimen3\font minus
  \fontdimen4\font\relax}
\providecommand{\BIBforeignlanguage}[2]{{%
\expandafter\ifx\csname l@#1\endcsname\relax
\typeout{** WARNING: IEEEtran.bst: No hyphenation pattern has been}%
\typeout{** loaded for the language `#1'. Using the pattern for}%
\typeout{** the default language instead.}%
\else
\language=\csname l@#1\endcsname
\fi
#2}}
\providecommand{\BIBdecl}{\relax}
\BIBdecl

\bibitem{borkar-dac-2007}
S.~Borkar, ``{Thousand Core Chips: A Technology Perspective},'' in
  \emph{Proceedings of 44th Annual Design Automation Conference (DAC)}.\hskip
  1em plus 0.5em minus 0.4em\relax New York, NY, USA: ACM, 2007, pp. 746--749.

\bibitem{kurian-ics-2014}
G.~Kurian, S.~Devadas, and O.~Khan, ``{Locality-aware Data Replication in the
  Last-level Cache},'' in \emph{Proceedings of 20th International Symposium on
  High Performance Computer Architecture (HPCA)}.\hskip 1em plus 0.5em minus
  0.4em\relax IEEE, 2014, pp. 1--12.

\bibitem{chen_dac_2016}
X.~Chen, N.~Khoshavi, J.~Zhou, D.~Huang, R.~DeMara, J.~Wang, W.~Wen, and
  Y.~Chen, ``{AOS: Adaptive Overwrite Scheme for Energy Efﬁcient MLC STT-RAM
  Cache},'' in \emph{Proceedings of 53nd Annual Design Automation Conference
  (DAC)}.\hskip 1em plus 0.5em minus 0.4em\relax ACM, 2016.

\bibitem{ahmadian-2014}
M.~Ahmadian, A.~Paya, and C.~D. Marinescu, ``{Security of Applications
  Involving Multiple Organizations - Order Preserving Encryption in Hybrid
  Cloud Environments},'' in \emph{Proceedings of 28th International Parallel
  and Distributed Processing Symposium Workshops}.\hskip 1em plus 0.5em minus
  0.4em\relax IEEE, 2014, pp. 894--903.

\bibitem{borkar-acm-2011}
S.~Borkar and A.~A. Chien, ``{The future of microprocessors},''
  \emph{Communications of the ACM}, vol.~54, no.~5, pp. 67--77, 2011.

\bibitem{TMSCS-memory}
D.~Kim, S.~Yoo, and S.~Lee, ``{Hybrid Main Memory for High Bandwidth Multi-Core
  System},'' \emph{IEEE Transactions on Multi-Scale Computing Systems}, vol.~1,
  no.~3, pp. 138--149, July 2015.

\bibitem{Imani-GLSVLSI-2016}
M.~Imani, S.~Patil, and T.~Rosing, ``{DCC: Double Capacity Cache for
  Narrow-Width Data Values},'' \emph{Great Lakes Symposium on VLSI}, 2016.

\bibitem{jog-dac-2012}
A.~Jog, A.~K. Mishra, C.~Xu, Y.~Xie, V.~Narayanan, R.~Iyer, and C.~R. Das,
  ``{Cache Revive: Architecting Volatile STT-RAM Caches for Enhanced
  Performance in CMPs},'' in \emph{Proceedings of 49th Annual Design Automation
  Conference (DAC)}.\hskip 1em plus 0.5em minus 0.4em\relax ACM, 2012, pp.
  243--252.

\bibitem{jaleel-hpca-2006}
A.~Jaleel, M.~Mattina, and B.~Jacob, ``{Last Level Cache (LLC) Performance of
  Data Mining Workloads on a CMP-a Case Study of Parallel Bioinformatics
  Workloads},'' in \emph{Proceedings of 12th International Symposium on High
  Performance Computer Architecture (HPCA)}.\hskip 1em plus 0.5em minus
  0.4em\relax IEEE, 2006, pp. 88--98.

\bibitem{Imani-NVMSA-2016}
M.~Imani, A.~Rahimi, Y.~Kim, and T.~Rosing, ``{A Low-Power Hybrid Magnetic
  Cache Architecture Exploiting Narrow-Width Values},'' \emph{Non-Volatile
  Memory Systems and Applications Symposium}, 2016.

\bibitem{bojnordi2013desc}
M.~N. Bojnordi and E.~Ipek, ``{DESC: Energy-efficient Data Exchange using
  Synchronized Counters},'' in \emph{Proceedings of 46th Annual IEEE/ACM
  International Symposium on Microarchitecture}.\hskip 1em plus 0.5em minus
  0.4em\relax ACM, 2013, pp. 234--246.

\bibitem{udipi2009non}
A.~N. Udipi, N.~Muralimanohar, and R.~Balasubramonian, ``{Non-uniform Power
  Access in Large Caches with Low-swing Wires},'' in \emph{Proceedings of
  International Conference on High Performance Computing (HiPC)}.\hskip 1em
  plus 0.5em minus 0.4em\relax IEEE, 2009, pp. 59--68.

\bibitem{loh-micro-2011}
G.~H. Loh and M.~D. Hill, ``{Efficiently Enabling Conventional Block Sizes for
  Very Large Die-stacked DRAM Caches},'' in \emph{Proceedings of 44th Annual
  IEEE/ACM International Symposium on Microarchitecture}.\hskip 1em plus 0.5em
  minus 0.4em\relax ACM, 2011, pp. 454--464.

\bibitem{arroyo-rd-2011}
R.~Arroyo, R.~Harrington, S.~Hartman, and T.~Nguyen, ``Ibm power7 systems,''
  \emph{IBM Journal of Research and Development}, vol.~55, no.~3, pp. 2--1,
  2011.

\bibitem{chang-hpca-2013}
M.-T. Chang, P.~Rosenfeld, S.-L. Lu, and B.~Jacob, ``{Technology Comparison for
  Large Last-level Caches (L 3 Cs): Low-leakage SRAM, Low Write-energy STT-RAM,
  and Refresh-optimized eDRAM},'' in \emph{Proceedings of 19th International
  Symposium on High Performance Computer Architecture (HPCA)}.\hskip 1em plus
  0.5em minus 0.4em\relax IEEE, 2013, pp. 143--154.

\bibitem{smullen-hpca-2011}
C.~W. Smullen, V.~Mohan, A.~Nigam, S.~Gurumurthi, and M.~R. Stan, ``{Relaxing
  Non-volatility for Fast and Energy-efficient STT-RAM Caches},'' in
  \emph{Proceedings of 17th International Symposium on High Performance
  Computer Architecture (HPCA)}.\hskip 1em plus 0.5em minus 0.4em\relax IEEE,
  2011, pp. 50--61.

\bibitem{khan2014improving}
S.~Khan, A.~R. Alameldeen, C.~Wilkerson, O.~Mutluy, and D.~A. Jimenezz,
  ``{Improving Cache Performance using Read-write Partitioning},'' in
  \emph{Proceedings of 20th International Symposium on High Performance
  Computer Architecture (HPCA)}.\hskip 1em plus 0.5em minus 0.4em\relax IEEE,
  2014, pp. 452--463.

\bibitem{tuck2003initial}
N.~Tuck and D.~M. Tullsen, ``{Initial observations of the simultaneous
  multithreading Pentium 4 processor},'' in \emph{Proceedings of 12th
  International Conference on Parallel Architectures and Compilation
  Techniques}.\hskip 1em plus 0.5em minus 0.4em\relax IEEE, 2003, pp. 26--34.

\bibitem{valero2015design}
A.~Valero, J.~Sahuquillo, P.~Lopez, and J.~Duato, ``{Design of Hybrid
  Second-Level Caches},'' \emph{IEEE Transactions on Computers}, vol.~64,
  no.~7, pp. 1884--1897, 2015.

\bibitem{agrawal2013refrint}
A.~Agrawal, P.~Jain, A.~Ansari, and J.~Torrellas, ``{Refrint: Intelligent
  Refresh to Minimize Power in on-chip Multiprocessor Cache Hierarchies},'' in
  \emph{Proceedings of 19th International Symposium on High Performance
  Computer Architecture (HPCA)}.\hskip 1em plus 0.5em minus 0.4em\relax IEEE,
  2013, pp. 400--411.

\bibitem{samavatian2014efficient}
M.~H. Samavatian, H.~Abbasitabar, M.~Arjomand, and H.~Sarbazi-Azad, ``{An
  efficient STT-RAM last level cache architecture for GPUs},'' in
  \emph{Proceedings of 51st ACM/EDAC/IEEE Design Automation Conference
  (DAC)}.\hskip 1em plus 0.5em minus 0.4em\relax IEEE, 2014, pp. 1--6.

\bibitem{sun-micro-2011}
Z.~Sun, X.~Bi, H.~H. Li, W.-F. Wong, Z.-L. Ong, X.~Zhu, and W.~Wu, ``{Multi
  Retention Level STT-RAM Cache Designs with a Dynamic Refresh Scheme},'' in
  \emph{Proceedings of 44th annual IEEE/ACM International Symposium on
  Microarchitecture}.\hskip 1em plus 0.5em minus 0.4em\relax ACM, 2011, pp.
  329--338.

\bibitem{pekhimenko2015exploiting}
G.~Pekhimenko, T.~Huberty, R.~Cai, O.~Mutlu, P.~B. Gibbons, M.~A. Kozuch, and
  T.~C. Mowry, ``{Exploiting Compressed Block Size as an Indicator of Future
  Reuse},'' in \emph{Proceedings of 21st International Symposium on igh
  Performance Computer Architecture (HPCA)}.\hskip 1em plus 0.5em minus
  0.4em\relax IEEE, 2015, pp. 51--63.

\bibitem{seshadri2015mitigating}
V.~Seshadri, S.~Yedkar, H.~Xin, O.~Mutlu, P.~B. Gibbons, M.~A. Kozuch, and
  T.~C. Mowry, ``{Mitigating {Prefetcher-caused} {Pollution} using {Informed}
  {Caching} {Policies} for {Prefetched} {Blocks}},'' \emph{ACM Transactions on
  Architecture and Code Optimization (TACO)}, vol.~11, no.~4, p.~51, 2015.

\bibitem{TMSCS-noninclusive}
H.~Xu, Y.~Alkabani, R.~Melhem, and A.~Jones, ``{FusedCache: A Naturally
  Inclusive, Racetrack Memory, Dual-Level Private Cache},'' \emph{IEEE
  Transactions on Multi-Scale Computing Systems}, vol.~PP, no.~99, pp. 1--1,
  2016.

\bibitem{zhao2012failure}
W.~Zhao, Y.~Zhang, T.~Devolder, J.-O. Klein, D.~Ravelosona, C.~Chappert, and
  P.~Mazoyer, ``{Failure and Reliability Analysis of STT-MRAM},''
  \emph{Microelectronics Reliability}, vol.~52, no.~9, pp. 1848--1852, 2012.

\bibitem{jiang2016improving}
L.~Jiang, W.~Wen, D.~Wang, and L.~Duan, ``{Improving Read Performance of
  STT-MRAM based Main Memories through Smash Read and Flexible Read},'' in
  \emph{Proceedings of 21st Asia and South Pacific Design Automation Conference
  (ASP-DAC)}.\hskip 1em plus 0.5em minus 0.4em\relax IEEE, 2016, pp. 31--36.

\bibitem{eken2014novel}
E.~Eken, Y.~Zhang, W.~Wen, R.~Joshi, H.~Li, and Y.~Chen, ``{A Novel
  Self-Reference Technique for STT-RAM Read and Write Reliability
  Enhancement},'' \emph{IEEE Transactions on Magnetics}, vol.~50, no.~11, pp.
  1--4, 2014.

\bibitem{li2010design}
J.~Li, P.~Ndai, A.~Goel, S.~Salahuddin, and K.~Roy, ``{Design Paradigm for
  Robust Spin-Torque Transfer Magnetic RAM (STT MRAM) From Circuit/Architecture
  Perspective},'' \emph{IEEE Transactions on Very Large Scale Integration
  (VLSI) Systems}, vol.~18, no.~12, pp. 1710--1723, 2010.

\bibitem{jokarsequoia}
M.~R. Jokar, M.~Arjomand, and H.~Sarbazi-Azad, ``{Sequoia: A High-Endurance
  NVM-Based Cache Architecture},'' \emph{IEEE Transactions on Very Large Scale
  Integration (VLSI) Systems}, 2016.

\bibitem{TMSCS-wear-leveling}
C.~Pan, S.~Gu, M.~Xie, C.~Xue, and J.~Hu, ``{Wear-Leveling Aware Page
  Management for Non-Volatile Main Memory on Embedded Systems},'' \emph{IEEE
  Transactions on Multi-Scale Computing Systems}, vol.~PP, no.~99, pp. 1--1,
  2016.

\bibitem{kang-aspdac-2014}
W.~Kang, W.~Zhao, Z.~Wang, J.-O. Klein, Y.~Zhang, D.~Chabi, Y.~Zhang,
  D.~Ravelosona, and C.~Chappert, ``{An Overview of Spin-based Integrated
  Circuits},'' in \emph{Proceedings of 19th Asia and South Pacific Design
  Automation Conference (ASP-DAC)}.\hskip 1em plus 0.5em minus 0.4em\relax
  IEEE, 2014, pp. 676--683.

\bibitem{worledge2011spin}
D.~Worledge, G.~Hu, D.~W. Abraham, J.~Sun, P.~Trouilloud, J.~Nowak, S.~Brown,
  M.~Gaidis, E.~O'Sullivan, and R.~Robertazzi, ``{Spin Torque Switching of
  Perpendicular Ta form},'' \emph{Applied Physics Letters}, vol.~98, no.~2, p.
  2501, 2011.

\bibitem{dong-tcad-2012}
X.~Dong, C.~Xu, Y.~Xie, and N.~P. Jouppi, ``{Nvsim: A Circuit-level
  Performance, Energy, and Area Model for Emerging Nonvolatile Memory},''
  \emph{Computer-Aided Design of Integrated Circuits and Systems, IEEE
  Transactions on}, vol.~31, no.~7, pp. 994--1007, 2012.

\bibitem{mars}
A.~Patel, F.~Afram, and K.~Ghose, ``{Marss-x86: A qemu-based
  micro-architectural and systems simulator for x86 multicore processors},'' in
  \emph{1st International Qemu Users’ Forum}, 2011, pp. 29--30.

\bibitem{Poremba-date-2015}
M.~Poremba, S.~Mittal, D.~Li, J.~S. Vetter, and Y.~Xie, ``{DESTINY: A Tool for
  Modeling Emerging 3D NVM and eDRAM caches},'' in \emph{Proceedings of 2015
  Design, Automation \& Test in Europe Conference \& Exhibition}.\hskip 1em
  plus 0.5em minus 0.4em\relax EDA Consortium, 2015, pp. 1543--1546.

\bibitem{khoshavi_isqed_2016}
N.~Khoshavi, X.~Chen, J.~Wang, and R.~DeMara, ``{Bit-Upset Vulnerability Factor
  for eDRAM Last Level Cache Immunity Analysis},'' in \emph{Proceedings of 17th
  International Symposium on Quality Electronic Design (ISQED)}.\hskip 1em plus
  0.5em minus 0.4em\relax IEEE, 2016.

\bibitem{Imani2016}
M.~Imani, S.~Patil, and T.~Rosing, ``{Low Power Data-Aware STT-RAM based Hybrid
  Cache Architecture},'' \emph{Proceedings of 17th International Symposium on
  Quality Electronic Design (ISQED)}, 2016.

\bibitem{wang2014adaptive}
Z.~Wang, D.~A. Jim{\'e}nez, C.~Xu, G.~Sun, and Y.~Xie, ``{Adaptive Placement
  and Migration Policy for an STT-RAM-based Hybrid Cache},'' in
  \emph{Proceedings of 20th International Symposium on High Performance
  Computer Architecture (HPCA)}.\hskip 1em plus 0.5em minus 0.4em\relax IEEE,
  2014, pp. 13--24.

\end{thebibliography}
\end{document}